\documentclass[aps,prl,reprint,amsmath,amssymb,longbibliography]{revtex4-2}

\ifdefined\pdfglyphtounicode
  \pdfglyphtounicode{parenleftbig}{0028 FE01}
  \pdfglyphtounicode{parenrightbig}{0029 FE01}
  \pdfglyphtounicode{bracketleftbig}{005B FE01}
  \pdfglyphtounicode{bracketrightbig}{005D FE01}
  \pdfglyphtounicode{floorleftbig}{230A FE01}
  \pdfglyphtounicode{floorrightbig}{230B FE01}
  \pdfglyphtounicode{ceilingleftbig}{2308 FE01}
  \pdfglyphtounicode{ceilingrightbig}{2309 FE01}
  \pdfglyphtounicode{braceleftbig}{007B FE01}
  \pdfglyphtounicode{bracerightbig}{007D FE01}
  \pdfglyphtounicode{angbracketleftbig}{27E8 FE01}
  \pdfglyphtounicode{angbracketrightbig}{27E9 FE01}
  \pdfglyphtounicode{vextendsingle}{20D3}
  \pdfglyphtounicode{vextenddouble}{20E6}
  \pdfglyphtounicode{slashbig}{002F FE01}
  \pdfglyphtounicode{backslashbig}{005C FE01}
  \pdfglyphtounicode{parenleftBig}{0028 FE02}
  \pdfglyphtounicode{parenrightBig}{0029 FE02}
  \pdfglyphtounicode{parenleftbigg}{0028 FE03}
  \pdfglyphtounicode{parenrightbigg}{0029 FE03}
  \pdfglyphtounicode{bracketleftbigg}{005B FE03}
  \pdfglyphtounicode{bracketrightbigg}{005D FE03}
  \pdfglyphtounicode{floorleftbigg}{230A FE03}
  \pdfglyphtounicode{floorrightbigg}{230B FE03}
  \pdfglyphtounicode{ceilingleftbigg}{2308 FE03}
  \pdfglyphtounicode{ceilingrightbigg}{2309 FE03}
  \pdfglyphtounicode{braceleftbigg}{007B FE03}
  \pdfglyphtounicode{bracerightbigg}{007D FE03}
  \pdfglyphtounicode{angbracketleftbigg}{27E8 FE03}
  \pdfglyphtounicode{angbracketrightbigg}{27E9 FE03}
  \pdfglyphtounicode{slashbigg}{002F FE03}
  \pdfglyphtounicode{backslashbigg}{005C FE03}
  \pdfglyphtounicode{parenleftBigg}{0028 FE04}
  \pdfglyphtounicode{parenrightBigg}{0029 FE04}
  \pdfglyphtounicode{bracketleftBigg}{005B FE04}
  \pdfglyphtounicode{bracketrightBigg}{005D FE04}
  \pdfglyphtounicode{floorleftBigg}{230A FE04}
  \pdfglyphtounicode{floorrightBigg}{230B FE04}
  \pdfglyphtounicode{ceilingleftBigg}{2308 FE04}
  \pdfglyphtounicode{ceilingrightBigg}{2309 FE04}
  \pdfglyphtounicode{braceleftBigg}{007B FE04}
  \pdfglyphtounicode{bracerightBigg}{007D FE04}
  \pdfglyphtounicode{angbracketleftBigg}{27E8 FE04}
  \pdfglyphtounicode{angbracketrightBigg}{27E9 FE04}
  \pdfglyphtounicode{slashBigg}{002F FE04}
  \pdfglyphtounicode{backslashBigg}{005C FE04}
  \pdfglyphtounicode{slashBig}{002F FE02}
  \pdfglyphtounicode{backslashBig}{005C FE02}
  \pdfglyphtounicode{parenlefttp}{239B}
  \pdfglyphtounicode{parenrighttp}{239E}
  \pdfglyphtounicode{bracketlefttp}{23A1}
  \pdfglyphtounicode{bracketrighttp}{23A4}
  \pdfglyphtounicode{bracketleftbt}{23A3}
  \pdfglyphtounicode{bracketrightbt}{23A6}
  \pdfglyphtounicode{bracketleftex}{23A2}
  \pdfglyphtounicode{bracketrightex}{23A5}
  \pdfglyphtounicode{bracelefttp}{23A7}
  \pdfglyphtounicode{bracerighttp}{23AB}
  \pdfglyphtounicode{braceleftbt}{23A9}
  \pdfglyphtounicode{bracerightbt}{23AD}
  \pdfglyphtounicode{braceleftmid}{23A8}
  \pdfglyphtounicode{bracerightmid}{23AC}
  \pdfglyphtounicode{braceex}{23AA}
  \pdfglyphtounicode{arrowvertex}{23D0}
  \pdfglyphtounicode{parenleftbt}{239D}
  \pdfglyphtounicode{parenrightbt}{23A0}
  \pdfglyphtounicode{parenleftex}{239C}
  \pdfglyphtounicode{parenrightex}{239F}
  \pdfglyphtounicode{angbracketleftBig}{27E8 FE02}
  \pdfglyphtounicode{angbracketrightBig}{27E9 FE02}
  \pdfglyphtounicode{unionsqtext}{2A06 FE01}
  \pdfglyphtounicode{unionsqdisplay}{2A06 FE02}
  \pdfglyphtounicode{contintegraltext}{222E FE01}
  \pdfglyphtounicode{contintegraldisplay}{222E FE02}
  \pdfglyphtounicode{circledottext}{2A00 FE01}
  \pdfglyphtounicode{circledotdisplay}{2A00 FE02}
  \pdfglyphtounicode{circleplustext}{2A01 FE01}
  \pdfglyphtounicode{circleplusdisplay}{2A01 FE02}
  \pdfglyphtounicode{circlemultiplytext}{2A02 FE01}
  \pdfglyphtounicode{circlemultiplydisplay}{2A02 FE02}
  \pdfglyphtounicode{summationtext}{2211 FE01}
  \pdfglyphtounicode{producttext}{220F FE01}
  \pdfglyphtounicode{integraltext}{222B FE01}
  \pdfglyphtounicode{uniontext}{22C3 FE01}
  \pdfglyphtounicode{intersectiontext}{22C2 FE01}
  \pdfglyphtounicode{unionmultitext}{2A04 FE01}
  \pdfglyphtounicode{logicalandtext}{22C0 FE01}
  \pdfglyphtounicode{logicalortext}{22C1 FE01}
  \pdfglyphtounicode{summationdisplay}{2211 FE02}
  \pdfglyphtounicode{productdisplay}{220F FE02}
  \pdfglyphtounicode{integraldisplay}{222B FE02}
  \pdfglyphtounicode{uniondisplay}{22C3 FE02}
  \pdfglyphtounicode{intersectiondisplay}{22C2 FE02}
  \pdfglyphtounicode{unionmultidisplay}{2A04 FE02}
  \pdfglyphtounicode{logicalanddisplay}{22C0 FE02}
  \pdfglyphtounicode{logicalordisplay}{22C1 FE02}
  \pdfglyphtounicode{coproducttext}{2210 FE01}
  \pdfglyphtounicode{coproductdisplay}{2210 FE02}
  \pdfglyphtounicode{hatwide}{02C6 FE01}
  \pdfglyphtounicode{hatwider}{02C6 FE02}
  \pdfglyphtounicode{hatwidest}{02C6 FE03}
  \pdfglyphtounicode{tildewide}{02DC FE01}
  \pdfglyphtounicode{tildewider}{02DC FE02}
  \pdfglyphtounicode{tildewidest}{02DC FE03}
  \pdfglyphtounicode{bracketleftBig}{005B FE02}
  \pdfglyphtounicode{bracketrightBig}{005D FE02}
  \pdfglyphtounicode{floorleftBig}{230A FE02}
  \pdfglyphtounicode{floorrightBig}{230B FE02}
  \pdfglyphtounicode{ceilingleftBig}{2308 FE02}
  \pdfglyphtounicode{ceilingrightBig}{2309 FE02}
  \pdfglyphtounicode{braceleftBig}{007B FE02}
  \pdfglyphtounicode{bracerightBig}{007D FE02}
  \pdfglyphtounicode{radicalbig}{221A FE01}
  \pdfglyphtounicode{radicalBig}{221A FE02}
  \pdfglyphtounicode{radicalbigg}{221A FE03}
  \pdfglyphtounicode{radicalBigg}{221A FE04}
  \pdfglyphtounicode{radicalbt}{23B7}
  \pdfglyphtounicode{radicalvertex}{20D3}
  \pdfglyphtounicode{radicaltp}{231C}
  \pdfglyphtounicode{arrowvertexdbl}{20E6}
  \pdfglyphtounicode{arrowtp}{2191}
  \pdfglyphtounicode{arrowbt}{2193}
  \pdfglyphtounicode{bracehtipdownleft}{23DF}
  \pdfglyphtounicode{bracehtipdownright}{23DF}
  \pdfglyphtounicode{bracehtipupleft}{23DE}
  \pdfglyphtounicode{bracehtipupright}{23DE}
  \pdfglyphtounicode{arrowdbltp}{21D1}
  \pdfglyphtounicode{arrowdblbt}{21D3}
\fi

\usepackage[T1]{fontenc}
\usepackage{lmodern,graphicx,bm,mathtools,microtype,xcolor}
\definecolor{linkblue}{HTML}{173D56}
\usepackage[colorlinks=true,allcolors=linkblue]{hyperref}
\hypersetup{pdftitle={Universal Critical Relations and Nonuniversal Scaling in Inverse-Square Ising Chains},pdfauthor={Dyuman Bhattacharya}}

\newcommand{\dd}{\mathrm d}

\DeclareMathOperator{\Var}{Var}

\begin{document}

\nocite{Thouless,ACCN,Jump,AYH,Kosterlitz,Bhattacharjee1981,Cardy1981,LuijtenMessingfeld,SM,Benedetti2026,DT,Wendland,GKS,Forest,Georgii,Ueltschi,FKG,Clark1987,Lebowitz}

\title{Universal Critical Relations and Nonuniversal Scaling in Inverse-Square Ising Chains}
\author{Dyuman Bhattacharya}
\email{d7bhatta@uwaterloo.ca}
\affiliation{Department of Physics and Astronomy, University of Waterloo, Waterloo, Ontario, Canada}
\date{November 8, 2026}
\begin{abstract}
For ferromagnetic Ising chains with $n^2J_n\to B>0$, the critical
excess magnetization in a field inversely proportional to spin
separation determines the connected correlation. Their slowly varying profile has finite logarithmic
integral exactly when the ordered thermal slope is finite. For any
finite nonnegative core, an exact tail gives inverse-logarithmic
correlations and a square-root cusp. A relative correction
$c/\log(n+e)$ with $c>1$ gives continuously varying logarithmic
powers and a linear cusp. A lower bound on its disordered susceptibility
excludes the usual Kosterlitz--Thouless law.
\end{abstract}
\maketitle

For the inverse-square Ising ferromagnet the spontaneous magnetization
is already positive at the transition~\cite{Thouless,ACCN}. Whenever
$J_n\sim B/n^2$ with nonnegative couplings, it satisfies
$2\beta_cBm_c^2=1$~\cite{Jump}. Once the tail strength and critical
temperature are known, this fixes the critical plateau. How do the
two-spin expectation and the magnetization in a small positive field
approach their respective plateaus?

The exact inverse-square chain is a standard model of
Kosterlitz--Thouless criticality~\cite{AYH,Kosterlitz,Bhattacharjee1981,Cardy1981,LuijtenMessingfeld}.
A tail correction may tend to zero in relative size at large
distances and still have a divergent sum over scales. We will see
how it can change a square-root cusp into a linear one. In both
cases the approach to the plateau in a field determines the decay
of the connected correlation and the finiteness of the thermal slope.
The finite nonnegative core is arbitrary.

\emph{A common physical critical profile.---}
Consider spins $\sigma_i=\pm1$ with Gibbs exponent
\begin{equation}\label{main:model}
 \begin{gathered}
 \beta\sum_{i<j}J_{j-i}\sigma_i\sigma_j+h\sum_i\sigma_i,\\
 0\le J_n<\infty,\qquad n^2J_n\to B>0.
 \end{gathered}
\end{equation}
Each bond occurs once and $h$ is dimensionless. The interaction is
fixed, including in constants and preparation scales. No convergence
rate, monotonicity, summable tail defect, reflection positivity or
positive-mixture representation is required.

Write $P_\beta(h)$ for the pressure per site,
$m_\beta(h)=P_\beta'(h)$ for $h>0$, and $m_\beta=P_\beta'(0+)$.
At $\beta_c=\inf\{\beta:m_\beta>0\}$ let $m_c=m_{\beta_c}>0$,
$P_c=P_{\beta_c}$. For the maximal plus state ($+$) write
\begin{equation}\label{main:profile}
 \begin{split}
 G_c(r)&=\langle\sigma_0\sigma_r\rangle_{\beta_c}^{+}-m_c^2,\\
 S(R)&=\frac{2}{m_c}[m_{\beta_c}(1/R)-m_c],\qquad R>0.
 \end{split}
\end{equation}
The tail hypothesis gives $S>0$, $S(R)\to0$, and
$S(aR)/S(R)\to1$ for every fixed $a>0$. Thus $R^\alpha S(R)\to\infty$
for every $\alpha>0$. Through integer separations $r\to\infty$,
\begin{equation}\label{main:ratio}
 \boxed{\displaystyle
 \frac{G_c(r)}{2m_c[m_{\beta_c}(a/r)-m_c]}\longrightarrow1,
 \qquad a>0.}
\end{equation}
with $m_{\beta_c}(a/r)+m_c\to2m_c$ supplying the coefficient in the
squared-field comparison (End Matter). Equation~\eqref{main:ratio} at
$a=1$ gives $G_c(r)\sim m_c^2S(r)$; rescaling the field replaces
$S(r)$ by $S(r/a)\sim S(r)$. For a block of length $N$, with mean
$\bar\sigma_N=N^{-1}\sum_{x=1}^N\sigma_x$ and truncated susceptibility
$\chi_c^{\rm tr}(N)=\sum_{|r|\le N}G_c(r)$,
\begin{equation}\label{main:collective}
 \Var_c^+(\bar\sigma_N)\sim
 \frac{\chi_c^{\rm tr}(N)}{2N}\sim m_c^2S(N).
\end{equation}
In the variance times $N$, the diagonal gives $G_c(0)=1-m_c^2$,
the off-diagonal $2\sum_{r=1}^{N-1}(1-r/N)G_c(r)$. There are
$N-r$ pairs at separation $r$. Slow variation gives $NG_c(N)/2$
for the triangular sum and $NG_c(N)$ for the unweighted sum.

Integrating the excess gives $(m_c/2)\int_0^h S(1/t)\,\dd t$, with
$\int_0^h S(1/t)\,\dd t\sim hS(1/h)$. The source estimates give
\begin{align}
 P_c(h)-P_c(0)-m_ch&\sim\frac{m_c}{2}hS(1/h),
 \label{main:pressure}\\
 0\le h m_{\beta_c}'(h)&=o(S(1/h)).
 \label{main:differential}
\end{align}
The little-o bound relative to the secant slope leaves the tangent
asymptotic open. End Matter bounds the squared-field error (proofs
in Supplemental Material~\cite{SM}).
In the magnetized state the block variance decays more slowly than
every inverse power of length, and
\begin{equation}\label{main:critical-growth}
 \frac{\log\chi_c^{\rm tr}(N)}{\log N}\longrightarrow1,
 \qquad
 \frac{m_{\beta_c}(h)-m_c}{h}\longrightarrow\infty.
\end{equation}
The connected sum grows almost linearly for every such profile.

\emph{Critical fluctuations classify the ordered thermal slope.---}
For fixed sufficiently large $R_0$, let
\begin{equation}\label{main:integrability}
 \mathcal I_J=\int_{R_0}^{\infty}S(R)\frac{\dd R}{R}.
\end{equation}
The right derivative of $m_\beta$ at $\beta_c$ exists in
$(0,\infty]$. It is finite precisely when $\mathcal I_J<\infty$.
Putting $h=1/R$ gives the field test below, while summing the
correlation over integer distances gives the spatial test. For
any sufficiently small $h_0>0$,
\begin{equation}\label{main:equivalent-tests}
 \int_0^{h_0}[m_{\beta_c}(h)-m_c]\frac{\dd h}{h}<\infty
 \quad\Longleftrightarrow\quad
 \sum_{r\ge1}\frac{G_c(r)}r<\infty .
\end{equation}
Neither test requires an expansion of the tail. Under the measure
$\dd R/R$, the band $[2^j,2^{j+1}]$ contributes
$(\log 2)S(2^j)[1+o(1)]$ by slow variation. The sum diverges for
the exact tail and converges for the logarithmically corrected
family below.

At the physical inverse temperatures $\beta_c\pm\tau$, $\tau>0$,
write the ordered excess as
\begin{equation}\label{main:thermal-defs}
 E_+(\tau)=2(\beta_c+\tau)B m_{\beta_c+\tau}^{2}-1.
\end{equation}
For $\mathcal I_J<\infty$ there is an intrinsic $\kappa_J>1/\beta_c$
such that
\begin{equation}\label{main:finite-slope}
 \begin{split}
 E_+(\tau)&=\kappa_J\tau+o(\tau),\\
 m_{\beta_c+\tau}-m_c
 &=\frac{m_c}{2}(\kappa_J-\beta_c^{-1})\tau+o(\tau).
 \end{split}
\end{equation}
For $\mathcal I_J=\infty$ one has $E_+(\tau)/\tau\to\infty$ and
$m_{\beta_c+\tau}-m_c\sim m_cE_+(\tau)/2$. Only the divergence of the slope is determined here. The power of the
cusp requires more information about the tail. For the finite slope
in Eq.~\eqref{main:finite-slope} the subtraction follows by division
by the critical jump identity,
\begin{equation}\label{main:ward-cusp}
 m_{\beta_c+\tau}-m_c
 =m_c\left[\sqrt{\frac{1+E_+(\tau)}{1+\tau/\beta_c}}-1\right].
\end{equation}
For a finite slope, the denominator subtracts $\tau/(2\beta_c)$
from the relative magnetization increment. This is the contribution
of the explicit inverse temperature in $2\beta Bm_\beta^2$.

\emph{Two microscopic regimes.---}
Suppose $J_n=B/n^2$ beyond any fixed finite nonnegative core. Then
\begin{equation}\label{main:exact-spatial}
 G_c(r)\sim\frac{m_c^2}{\log r},
 \qquad r\in\mathbb N.
\end{equation}
Writing $\ell_h=\log(1/h)$, the field and pressure laws are
\begin{align}
 m_{\beta_c}(h)-m_c
 &\sim\frac{m_c}{2\ell_h},
 \label{main:exact-field}\\
 P_c(h)-P_c(0)-m_ch
 &\sim\frac{m_ch}{2\ell_h}.
 \label{main:exact-pressure}
\end{align}
Equation~\eqref{main:collective} gives $m_c^2/\log N$ for both
collective quantities. The leading amplitudes are expressed
in terms of $m_c$. The finite core enters the error estimates
(Supplemental Sec.~S5) and the thermal metric below. The truncated
flow of Ref.~\cite{Benedetti2026} also gives the inverse logarithm;
Sec.~S5 bounds the error for the original spins at all integer
separations. Since $S(R)\sim1/\log R$, its integral up to $X$
is asymptotic to $\log\log X$ as $X\to\infty$. For this tail
\begin{equation}\label{main:exact-metric}
 \lambda_J=\lim_{\tau\downarrow0}\frac{E_+(\tau)^2}{\tau}
 \quad\hbox{exists in }(0,\infty),
\end{equation}
and
\begin{equation}\label{main:exact-cusp}
 E_+(\tau)\sim\sqrt{\lambda_J\tau},\qquad
 m_{\beta_c+\tau}-m_c\sim\frac{m_c}{2}\sqrt{\lambda_J\tau}.
\end{equation}
The metric of the given interaction refers to no preparation scale
or reduced coordinates. The pure chain is included.

Keep an arbitrary finite nonnegative core and now put
\begin{equation}\label{main:log-tail}
 J_n=\frac{B}{n^2}\left(1+\frac{c}{\log(n+e)}\right),
 \qquad c>1.
\end{equation}
This vanishing relative correction gives
\begin{equation}\label{main:log-profile}
 S(R)\sim A_J(\log R)^{1-2c},\qquad E_+(\tau)\sim\kappa_J\tau,
\end{equation}
where $A_J>0$ and $\kappa_J>1/\beta_c$. Putting $t=\log R$ gives
$\int^\infty t^{1-2c}\,\dd t<\infty$, since $1-2c<-1$.
The linear cusp has coefficient given by Eq.~\eqref{main:finite-slope}.
Varying $c>1$ realizes every logarithmic power $2c-1>1$ at fixed
leading tail and jump law. Equations~\eqref{main:ratio}--\eqref{main:pressure}
transfer these powers to the spatial and field observables. A finite-core
change preserves them; field or distance rescaling only shifts the
logarithm (Table~\ref{tab:regimes}).

An interval of $N$ reversed spins in a uniform background costs
\begin{equation}\label{main:droplet}
 \begin{split}
 \Delta E_J(N)&=4\sum_{n\ge1}\min(n,N)J_n\\
 &=4B\log N+4Bc\log\log N+O_J(1).
 \end{split}
\end{equation}
There are $2\min(n,N)$ crossing bonds of length $n$, each costing $2J_n$.
The energy cost enters the Gibbs weight multiplied by $\beta$. For
Eq.~\eqref{main:log-tail} the crossing bonds of length at most $N$
contribute a term proportional to $\sum_{2\le n\le N}1/(n\log n)$.
This gives the double logarithm in Eq.~\eqref{main:droplet}.
Obtaining a critical power requires the evolution of the wall activity.

\begin{table*}[t]
\caption{Critical and ordered observables for two tails beyond an
arbitrary finite nonnegative core. $A_J$, $\lambda_J$ and $\kappa_J$
depend on the interaction. The first two rows give leading equivalents.}
\label{tab:regimes}
\begin{ruledtabular}
\begin{tabular}{lcc}
 & Exact inverse-square tail & Logarithmic correction, $c>1$\\
\hline
$S(R)$ & $\displaystyle\frac1{\log R}$ & $A_J(\log R)^{1-2c}$\\
$E_+(\tau)$ & $\sqrt{\lambda_J\tau}$ & $\kappa_J\tau$\\
$\partial_\beta^+m_{\beta_c}$ & $+\infty$ & $\displaystyle\frac{m_c}{2}(\kappa_J-\beta_c^{-1})\in(0,\infty)$
\end{tabular}
\end{ruledtabular}
\end{table*}

\emph{A disordered consequence.---}
For $\beta<\beta_c$, $\chi_-(\beta)=P_\beta''(0)=\sum_{r\in\mathbb Z}
\langle\sigma_0\sigma_r\rangle_\beta$ is finite by sharpness~\cite{DT}.
The dimensionless field introduces no factor of $\beta$. This sum
extends over the full chain, whereas Eq.~\eqref{main:collective}
truncates the connected critical sum. For $\mathcal I_J<\infty$,
continuation on the disordered side yields
\begin{equation}\label{main:finite-matching}
 \tau\log\chi_-(\beta_c-\tau)\to\infty,
 \qquad E_+(\tau)\log\chi_-(\beta_c-\tau)\to\infty.
\end{equation}
Thus $\chi_-(\beta_c-\tau)$ grows faster than $\exp(C/\tau)$ for
any fixed $C>0$, excluding $\log\chi_-\sim C\tau^{-1/2}$.
The explicit family \eqref{main:log-tail} obeys
\begin{equation}\label{main:log-exit}
 \liminf_{\tau\downarrow0}
 \frac{\tau\log\chi_-(\beta_c-\tau)}{\log(1/\tau)}
 \ge\frac{c-1}{\kappa_J}.
\end{equation}
Since $E_+(\tau)\sim\kappa_J\tau$ at the same physical displacement,
replacing $\tau$ by $E_+(\tau)$ in the numerator gives the lower bound
$c-1$. For arbitrary tails
$\log\chi_-/\log(1/\tau)\to\infty$. The lower bound on the residual involving
the effective pressure curvature at the stopping scale gives the liminf above. An equivalent
would require upper control (End Matter and Supplemental Sec.~S4).

\emph{Physical origin of the relations.---}
Decompose the tail into Gaussian shells, retaining each covariance
as a function, its variance, and stable terms from earlier shells.
The compensated core is ferromagnetic. A two-state compression of
its transfer operator gives a finite preparation $H_0$ and block
factor $L$ for one $C^2$ physical temperature curve in fixed
coefficient spaces. They may depend on the full interaction. The
finite-volume identities retain physical endpoints, marked spins,
magnetic sources and every scalar the sources generate (End Matter
and Supplemental Secs.~S1--S2).

At lengths $H_j=H_0L^j$, put $b=\log L$. At criticality the source factor obeys
$Z_j^2/m_c^2-1=q_j/b$, where $q_j$ is the physical excess charge.
On the positive critical orbit
\begin{equation}\label{main:critical-signal}
 q_j\sim u_j\sim\sum_{k\ge j}v_k^2,\qquad v_j^2=o(q_j),
\end{equation}
with reduced charge $u_j$ and wall activity $v_j$. The field gives
$m_{\beta_c}(h)=Z_j+O_J(v_j^2)$ for $h$ of order $H_j^{-1}$.
A pair gives $\langle\sigma_0\sigma_r\rangle_{\beta_c}^+=Z_j^2+O_J(v_j^2)$
for $r$ of order $H_j$. The two excesses are proportional because
$Z_j^2-m_c^2=(Z_j+m_c)(Z_j-m_c)$ with $Z_j+m_c\to2m_c$.
By Eq.~\eqref{main:critical-signal} the errors are smaller than the
excess however slowly the tail converges. A bounded depth shift
has vanishing relative effect, giving Eq.~\eqref{main:ratio} for
any fixed multiplier. In the marked-pair estimate take volume to
infinity at positive source. Removing the source with the separation
fixed selects the maximal plus state; then let the separation diverge.

The physical inverse-temperature curve crosses the critical separator
with strictly positive transverse derivative. Dividing out convergent
positive diagonal factors, the variations of charge and minus the
logarithm of activity satisfy
\begin{equation}\label{main:positive-variation}
 \begin{gathered}
 \binom{\widetilde U_{j+1}}{\widetilde W_{j+1}}=
 \begin{pmatrix}1&b_j\\c_j&1\end{pmatrix}
 \binom{\widetilde U_j}{\widetilde W_j},\\
 b_j\asymp v_j^2,\qquad c_j\asymp1.
 \end{gathered}
\end{equation}
Here $(\widetilde U_0,\widetilde W_0)=(1,0)$ varies charge at fixed
activity; $\asymp$ means upper and lower positive bounds uniform in
$j$. By the second row,
$c_*j\le\widetilde W_j\le C_*j\widetilde U_j$. The first row weights this response by a quantity of order $v_j^2$.
It bounds the charge increment below by a constant
times $jv_j^2$ and the ratio of successive charge responses above
by $1+Cjv_j^2$. Summing the lower bound or multiplying the upper
bounds gives finite gain precisely if $\sum_j(j+1)v_j^2<\infty$.
Indeed, summing Eq.~\eqref{main:critical-signal} counts each squared
activity $k+1$ times: $\sum_{j\ge0}\sum_{k\ge j}v_k^2
=\sum_{k\ge0}(k+1)v_k^2$. Thus $\sum_jq_j<\infty$ is the same test.
The depth sum becomes the integral in Eq.~\eqref{main:integrability}
on conversion to physical length. Supplemental Sec.~S4 controls the
nonlinear endpoint error in transferring the tangent gain to the
zero-field ordered slope and proves the disordered lower bounds.
Sec.~S5 gives the exact-tail square-root cusp.

The jump relation is the same in both examples. By
Eqs.~\eqref{main:ratio} and \eqref{main:integrability}, it is the
approach to that jump in a critical field that fixes the spatial
profile and decides the finiteness of the ordered thermal slope.
\phantomsection\label{main:core-end}

\paragraph*{Data and code availability.}
The Supplemental Material contains the derivations.

\bibliography{references}

\clearpage

\section*{End Matter}
\phantomsection\label{end:begin}
\setcounter{equation}{0}
\renewcommand{\theequation}{A\arabic{equation}}
\renewcommand{\theHequation}{endmatter.\arabic{equation}}

\paragraph*{A. Physical reconstruction.}
For each $J$ in Eq.~\eqref{main:model}, Supplemental Secs.~S1--S2
give one finite $H_0$, an integer $L$, and a $C^2$ physical temperature
curve on an open interval about $\beta_c$ in fixed coefficient spaces.
At $H_j=H_0L^j$ the prescribed covariance data $\omega_j$ satisfy
$\partial_\beta\omega_j=0$. The neutral and charged remainders have
norm $\|R\|=\|R_0\|+\|R_1\|$. We apply these maps at finite volume, with the endpoints free and the
two microscopic spin marks in place. We also keep each scalar generated
by the magnetic source when computing the pressure.

With $b=\log L$, the physical source factor $Z_j>0$, reduced charge
$u_j$ and wall activity $v_j>0$ obey
\begin{equation}\label{end:ward}
 \begin{split}
 p_j&=4\beta BZ_j^2,\qquad q_j=b(p_j/2-1),\\
 q_j&=u_j+O_J(v_j^2+\theta^j),\qquad 0<\theta<1.
 \end{split}
\end{equation}
and
\begin{equation}\label{end:map}
 \begin{split}
 u_{j+1}&=u_j-k_j(u_j,v_j)v_j^2,\\
 v_{j+1}&=v_j\exp[-u_j-(1+u_j/b)a_j+v_j^2h_j(u_j,v_j)].
 \end{split}
\end{equation}
The shell variance $C_j^{\rm sh}(0)$ gives the fixed defect
$a_j=C_j^{\rm sh}(0)-b\to0$. The coefficients depend on covariance
shape and earlier stable coordinates, with $k_j(0,0)\to1$ and
$|Dk_j|+|h_j|+|Dh_j|\le C$ for $D=D_{(u,v)}$.
Include every source-dependent scalar in the pressure $\widehat P_j$
per block. Its fixed-depth thermodynamic limit satisfies
\begin{equation}\label{end:source-identity}
 \begin{split}
 P_\beta(h_2)-P_\beta(h_1)=H_j^{-1}\bigl[&\widehat P_j(H_jZ_jh_2)\\
                         &-\widehat P_j(H_jZ_jh_1)\bigr].
 \end{split}
\end{equation}
At criticality, on a fixed positive interval of effective fields $\eta$,
\[
 \widehat P_j(\eta)=c_j^0+\eta+O_J(\epsilon_j),
 \qquad \epsilon_j=v_j^2+\theta^j,
\]
where $c_j^0$ is independent of $\eta$. Interior secants calibrate the
magnetization by $Z_j$. The selected critical orbit has $q_j\to0$ and
\begin{equation}\label{end:signal}
 \begin{gathered}
 u_j\sim\sum_{k\ge j}v_k^2,\qquad q_j\sim u_j,\\
 v_j^2+\theta^j=o(q_j),\qquad q_{j+1}/q_j\to1.
 \end{gathered}
\end{equation}
For large $R$ put $j(R)=\lfloor\log(R/H_0)/b\rfloor$. On each
distance band the field and spin estimates are
\begin{equation}\label{end:readouts}
 \begin{split}
 m_{\beta_c}(h)&=Z_{j(1/h)}+O_J(\epsilon_{j(1/h)}),\\
 \sup_{H_j\le r<H_{j+1}}\bigl|
 \langle\sigma_0\sigma_r\rangle_{\beta_c}^+-Z_j^2\bigr|
 &\le C_J\epsilon_j,\qquad r\in\mathbb N.
 \end{split}
\end{equation}
Combining these estimates at boundedly separated depths gives, for
each fixed $a>0$,
\begin{equation}\label{end:squared-field}
 G_c(r)=m_{\beta_c}(a/r)^2-m_c^2+O_{J,a}(v_{j(r)}^2).
\end{equation}
At $a=1$ the squared excess is $m_c^2[S(r)+S(r)^2/4]$.
Equation~\eqref{end:signal} gives $v_{j(r)}^2=o(S(r))$, proving
Eq.~\eqref{main:ratio}. Resolving the quadratic term would require
an $o(S(r)^2)$ error.

Differentiate both microscopic marks at entry. Once their supports
fit in a bounded window, the joint insertion splits as
\[
 \begin{gathered}
 E_j^{12}=A_j\mathbf1_j+P_{j,0}+P_{j,1},\qquad
 A_j=Z_j^2+O_J(v_j^2),\\
 \|P_{j,0}\|=O_J(v_j^2),\qquad \|P_{j,1}\|=O_J(v_j).
 \end{gathered}
\]
The complete unit $\mathbf1_j$ transports exactly. Only the proper
terms contract, and evaluation of the charged term $P_{j,1}$ gains
one factor of the activity. Together they contribute $O_J(v_j^2)$.
For fixed $r$, ferromagnetic monotonicity gives, along regular fields,
\[
 \langle\sigma_0\sigma_r\rangle_{\beta_c}^{+}
 =\lim_{h\downarrow0}\lim_{\Lambda\uparrow\mathbb Z}
       \langle\sigma_0\sigma_r\rangle_{\Lambda,\beta_c,h}.
\]
The large-$r$ estimate is taken in this state (Supplemental Sec.~S3).

\paragraph*{B. Thermal amplification and the physical branches.}
Write $w=\log v$. The critical separator is the graph
$u=\gamma_j(w)$, with slope $\ell_j\ge0$. Varying the physical
inverse temperature at entry displaces the initial point across
this graph with derivative
\[
 d_0=\partial_\beta[u_0-\gamma_0(\log v_0)]_{\beta_c}>0.
\]
Let $\Pi_j$ be the product of the complete normal multipliers through
scale $j$. The Riccati tangent equation and the positive charge sum give
\begin{equation}\label{end:gain}
 \begin{split}
 \Pi_\infty<\infty
 &\ \Longleftrightarrow\ \sum_j\ell_j<\infty
 \ \Longleftrightarrow\ \sum_j(j+1)v_j^2<\infty\\
 &\ \Longleftrightarrow\ \sum_jq_j<\infty
 \ \Longleftrightarrow\ \mathcal I_J<\infty.
 \end{split}
\end{equation}
The finite case has
$u_{\infty,+}(\tau)=d_0\Pi_\infty\tau+o(\tau)$; otherwise
$u_{\infty,+}(\tau)/\tau\to\infty$.
Uniform nonlinear difference bounds control the error in passing
from the finite-depth tangent to the ordered endpoint. Calibration
gives
$E_+=u_{\infty,+}/b$. Equation~\eqref{main:ward-cusp} then yields the
physical slope, with $\kappa_J=d_0\Pi_\infty/b>\beta_c^{-1}$ in
the finite case. A ferromagnetic lower bound on the magnetization
increment proves the strict inequality. Supplemental Sec.~S4 gives
these endpoint estimates.

The nonlinear estimate is uniform over nearby critical initial
activities. Start the ordered and critical trajectories at the same
activity, at
$(\gamma_0(w)+\delta,w)$ and $(\gamma_0(w),w)$, respectively.
Compare their charges by $d_j=u_j^+-u_j^c(w)$ and their activities
by $z_j=\log[v_j^c(w)/v_j^+]$. When the activity moment is finite,
the exact mean-value recurrences yield
\begin{equation}\label{end:nonlinear-endpoint}
 \begin{gathered}
 0\le d_j\le C\delta,\qquad
 0\le z_j\le C\delta(j+1),\\
 |u_{\infty,+}-d_N|
 \le C\delta\sum_{j\ge N}(j+1)[v_j^c(w)]^2.
 \end{gathered}
\end{equation}
The moment tails are uniformly dominated near the physical entry.
Consequently we can let $\delta\downarrow0$ at fixed $N$ and then
let $N\to\infty$. The uniform tail estimate identifies the derivative
of the endpoint with the limit of the finite-depth derivatives.
For the physical temperature curve the initial displacement is
$\delta=d_0\tau+o(\tau)$.

For an exact tail, the heat construction has geometric covariance
errors and
$u_j^c,v_j^c=j^{-1}+O_J(\log j/j^2)$.
The normal product satisfies $\Pi_j/j\to P_*>0$.
For $X_j=(u_j,v_j)$, its physical tangent has the calibrated growth
\begin{equation}\label{end:exact-tangent}
 \left.\partial_\beta X_j\right|_{\beta_c}
 =\frac{d_0P_*j}{3}(2,-1)+o(j).
\end{equation}
The map Hessian bound controls the nonlinear Taylor remainder by
$O_J(\tau^2j^3)$ when $\tau j^2$ is sufficiently small.
With $s_\tau=(d_0P_*\tau)^{-1/2}$, the rescaled ordered trajectory
$(U,V)=s_\tau(u_j,v_j)$ at time $t=j/s_\tau$ converges to
\begin{equation}\label{end:exact-ordered-flow}
 (U_+(t),V_+(t))
 =\sqrt2\bigl(\coth(\sqrt2t),\operatorname{csch}(\sqrt2t)\bigr),
\end{equation}
the solution of $U'=-V^2$, $V'=-UV$ with its calibrated incoming
tangent. An ordered charge-loss estimate transfers the endpoint:
$u_{\infty,+}\sim\sqrt2/s_\tau$ and
$\lambda_J=2d_0P_*/b^2>0$.
Supplemental Sec.~S5 derives this limit and the critical error
bounds with the same physical construction and source normalization.

For Eq.~\eqref{main:log-tail}, direct Fourier and heat-shell estimates
give $a_j=c/j+O_J(j^{-2})$.
The critical recurrence then yields, for $V_J>0$,
\begin{equation}\label{end:log-orbit}
 v_j^c\sim V_Jj^{-c},\qquad
 q_j^c\sim\frac{V_J^2}{2c-1}j^{1-2c}.
\end{equation}
Thus the moment in Eq.~\eqref{end:gain} converges. Conversion from
$j$ to physical length gives
$A_J=V_J^2b^{2c-2}/(2c-1)$ in Eq.~\eqref{main:log-profile}.
\paragraph*{C. Disordered stopping and susceptibility.}
Fix a sufficiently small $\varepsilon>0$ and a sufficiently late
integer $J_\varepsilon$, both independent of $\tau$. For the disordered
orbit at $\beta_c-\tau$, let
\[
 N_-(\tau)=\min\{j\ge J_\varepsilon:v_j^-(\tau)\ge\varepsilon\}.
\]
This first passage is finite and remains within the weak charts.
Changing between two fixed small sections changes $N_-$ by an amount
bounded independently of $\tau$; the bound may depend on both sections.
At $N=N_-(\tau)$, complete source curvature gives
\begin{equation}\label{end:complete-residual}
 \begin{split}
 \widehat\chi_N&=\widehat P_N''(0)
       =\frac{\chi_-(\beta_c-\tau)}{H_NZ_N^2},\\
 \Delta_-(\tau)&=\log(H_0Z_N^2\widehat\chi_N).
 \end{split}
\end{equation}
Consequently
\begin{equation}\label{main:residual}
 \begin{gathered}
 \log\chi_-(\beta_c-\tau)=bN_-(\tau)+\Delta_-(\tau),
 \\
 \Delta_-(\tau)\ge-C_{J,\varepsilon}.
 \end{gathered}
\end{equation}
Sharpness gives finite susceptibility at every fixed subcritical
temperature. The residual lower bound follows from comparison in a
positive field. Differentiating the pressure in
Eq.~\eqref{end:complete-residual} retains the scalar, mixed and
second source derivatives. Keeping only the variance of the final
spin weight would lose these contributions.

For every interaction in Eq.~\eqref{main:model},
$N_-(\tau)/\log(1/\tau)\to\infty$; the finite-slope class has
$\tau N_-(\tau)\to\infty$. For Eq.~\eqref{main:log-tail},
\begin{equation}\label{end:log-stopping}
 \log H_{N_-}\sim\frac{c-1}{\kappa_J\tau}\log\frac1\tau.
\end{equation}
The leading time is spent growing from activity of order $\tau^c$
to order $\tau$. Supplemental Sec.~S4 proves these passage laws.
Together with Eq.~\eqref{main:residual}, they yield the susceptibility
divergences and Eq.~\eqref{main:log-exit}. The residual has a lower
bound only. The stopping-scale equivalent therefore gives a lower
bound on susceptibility.

\phantomsection\label{end:finish}

\end{document}


\nocite{Thouless,ACCN,Jump,AYH,Kosterlitz,Bhattacharjee1981,Cardy1981,LuijtenMessingfeld,SM,Benedetti2026,DT,Wendland,GKS,Forest,Georgii,Ueltschi,FKG,Clark1987,Lebowitz}

\begin{center}
{\Large\bfseries Supplemental Material}\par
\vspace{5pt}
{\large Universal Critical Relations and Nonuniversal Scaling\\in Inverse-Square Ising Chains}\par
\vspace{7pt}
Dyuman Bhattacharya\par
{\small Department of Physics and Astronomy, University of Waterloo,\\Waterloo, Ontario, Canada}\par
{\small\href{mailto:d7bhatta@uwaterloo.ca}{d7bhatta@uwaterloo.ca}}
\end{center}
\medskip
\begingroup\footnotesize\setlength{\parskip}{0pt}\tableofcontents\endgroup

\section{Exact representation and common analytic construction}
\label{sec:representation-and-map}\label{sec:tail-representation}
\label{sec:common-maps}\label{sec:common-map}

\label{sec:setup}\label{sec:mandate}\label{s2:authority}
Fix a translation-invariant $J$ in
\begin{equation}
 \mathcal J_B=\{J:0\le J_n<\infty,\ n^2J_n\to B\},\qquad B>0.
 \label{s2:class}
\end{equation}
Constants and preparation may depend on $J$. For finite intervals
$\Lambda\subset\Z$, the free law on $\{-1,1\}^{\Lambda}$ and pressure are
\[
\begin{aligned}
 \mu_{\Lambda,h}(\sigma)&=\frac{2^{-|\Lambda|}}{Z_\Lambda(h)}
 \exp\!\left(\beta\sum_{\substack{i,j\in\Lambda\\ i<j}}J_{j-i}\sigma_i\sigma_j
       +h\sum_{i\in\Lambda}\sigma_i\right),\\
 P_\beta(h)&=\lim_{\Lambda\uparrow\Z}|\Lambda|^{-1}\log Z_\Lambda(h),\qquad
 m_\beta(h)=P_\beta'(h)\quad(h>0).
\end{aligned}
\]
The maximal plus state has $\langle F\rangle_\beta^+=
\lim_{h\downarrow0}\lim_{\Lambda\uparrow\Z}\mu_{\Lambda,h}(F)$ for local $F$.
Put $m_\beta=P_\beta'(0+)$, $\beta_c=\inf\{\beta:m_\beta>0\}$,
$m_c=m_{\beta_c}$, $P_c=P_{\beta_c}$ and $m_c(h)=m_{\beta_c}(h)$.
By the jump theorem~\cite{Jump},
\begin{equation}
 0<\beta_c<\infty,\qquad m_c>0,\qquad 2\beta_cBm_c^2=1.
 \label{new:jump-import}
\end{equation}
We use its diffuse-deletion, pressure-ceiling and logarithmic-energy
arguments and critical field lower bound (its Supplement, Secs.~S2--S3, S7).
We prove the finite preparation and response estimates below.

Put $G_c(r)=\langle\sigma_0\sigma_r\rangle_{\beta_c}^+-m_c^2$,
$\bar\sigma_N=N^{-1}\sum_{x=1}^N\sigma_x$ and
$\chi_c^{\rm tr}(N)=\sum_{|r|\le N}G_c(r)$, with $G_c(0)=1-m_c^2$.
With $R,\tau>0$ and physical detunings $\beta_c\pm\tau$, put
\begin{equation}
 S(R)=\frac2{m_c}[m_{\beta_c}(1/R)-m_c],\qquad
 E_+(\tau)=2(\beta_c+\tau)Bm_{\beta_c+\tau}^2-1.
 \label{new:physical-S}
\end{equation}
For $\beta<\beta_c$, sharpness~\cite[Theorem~2.1(2)]{DT} gives
$\chi_-(\beta)=P_\beta''(0)=\sum_{r\in\Z}\langle\sigma_0\sigma_r\rangle_\beta
<\infty$ without a factor of $\beta$ since $h$ is dimensionless.

At $H_j=H_0L^j$, $L\ge2$ an integer, put $b=\log L$. The source factor $Z_j>0$ gives
\begin{equation}
 p_j=4\beta BZ_j^2,\qquad q_j=b(p_j/2-1),\qquad
 h_{\rm eff}=H_jZ_jh.
 \label{new:Ward-conventions}
\end{equation}
The reduced charge $u_j$ and activity $v_j>0$ obey
$q_j=u_j+O_J(v_j^2+\theta^j)$, $0<\theta<1$.
The ordinary neutral and charged remainders $R_{0,j},R_{1,j}$ retain
source-dependent scalars. Physical endpoints and microscopic spins have
separate modules.

For the fixed interaction, put $j_n=J_n/B$ and
\begin{equation}
 \epsilon_n=n^2j_n-1,\qquad E(H)=\sup_{n\ge H}|\epsilon_n|\downarrow0.
 \label{s2:epsilon}
\end{equation}
The Gaussian field retains the full difference between $J$ and the
finite-range core. No rate for $E(H)$ is assumed.

\subsection{Sources, endpoints, and a positive finite core}
\label{s3:exact-source}\label{sec:canonical-positive-core}

For a free interval of $N$ spins put $q_0=\sigma_1/2$,
$q_i=(\sigma_{i+1}-\sigma_i)/2$ $(0<i<N)$, and $q_N=-\sigma_N/2$.
Then $\sum q_i=0$ and
$\sum q_i\varphi_i=-\frac12\sum_{i=1}^N\sigma_i(\varphi_i-\varphi_{i-1})$.
The half charges fix the endpoints and source sign. For the canonical
interaction $B/n^2$,
\begin{align}
 V_0(x)&=\int_0^\infty\frac{u(1-e^{-xu})}{4\sinh^2(u/2)}\,du
       =\log x+\gamma_E+1+r_0(x),\notag\\
 r_0(x)&=\int_0^\infty\left[\frac1u-\frac{u}{4\sinh^2(u/2)}\right]e^{-xu}\,du,
 &|r_0^{(j)}(x)|&\le\frac{(j+1)!}{12x^{j+2}}.
 \label{sm:ent:residual}
\end{align}
The bracket lies in $[0,u/12]$ by the partial fractions of $\coth(u/2)$;
$V_0(N)=\sum_{n\le N}n^{-1}+N\sum_{n>N}n^{-2}$. Let
\begin{equation}
 \phi(r)=(1-r)_+^9
 (1+9r+237r^2/7+453r^3/7+384r^4/7).
 \label{sm:ent:profile}
\end{equation}
With $\mathcal If(r)=\int_r^\infty t f(t)dt$ and $c_4=494208$,
$\phi=c_4\mathcal I^4[(1-r)_+^5]$ is positive definite on $\R$
by Ref.~\cite[Theorem~3.5]{Wendland}, with dimension $1$ and order $4$.
Define
\[
 \rho_\phi(s)=\frac{\int_s^\infty u^6(u^2-s^2)^3e^{-u}du}{48c_4\,5!},
 \qquad w_0(s)=\int_0^\infty\frac{u^2\rho_\phi(us)}{4\sinh^2(u/2)}du.
\]
The identity $e^{-x}=\int_0^\infty\rho_\phi(s)\phi(x/s)ds$ follows
by applying $\partial_x^6(-x^{-1}\partial_x)^4$: both sides give
$c_4\,5!\rho_\phi(x)x^{-13}$, and successive integration from infinity,
where all boundary terms vanish, recovers them. Taking $x\downarrow0$
gives $\int\rho_\phi=1$. Tonelli and $t=us$ give
\begin{equation}
 \begin{gathered}
 V_0(x)=\int_0^\infty w_0(s)[1-\phi(x/s)]ds,\\
 0<w_0(s)\le s^{-1},\qquad w_0'\le0,\qquad
 w_0(0)<\infty,\qquad w_0(s)=s^{-1}+O(s^{-3}).
 \end{gathered}
 \label{sm:ent:w}
\end{equation}
Here $\rho_\phi$ is positive, decreasing and has finite moments;
$u^2/[4\sinh^2(u/2)]\le1$ and equals $1+O(u^2)$ at zero.

For integer $H\ge1$, set $C^0_{<H}(x)=\int_0^Hw_0(s)\phi(|x|/s)ds$ and
$J_H^0(n)=B\Delta_d C^0_{<H}(n)$, where
$\Delta_df(n)=f(n+1)-2f(n)+f(n-1)$. These are positive covariance
pieces, and
\[
 (C^0_{<H})''(r)=-\frac{w_0(H)\phi'(r/H)}r
 +\int_{r/H}^1\frac{w_0'(r/u)\phi'(u)}{u^2}\,du\ge0.
\]
Since $\phi''<0$ on $[0,1/8]$ and $|\phi''|\le156/7$,
\begin{equation}
 J_H^0\ge0,\quad J_H^0(n)=0\ (n>H),\quad
 |J_H^0(n)-B/n^2|\le78B/(7H^2),\quad
 J_H^0(n)\ge B/n^2\ (n+1\le H/8).
 \label{sm:ent:corebounds}
\end{equation}
Moreover $J_H^0(n)\ge a_0B/H^2$ on $H/4\le n\le H/3$, for a fixed
$a_0>0$. Put $D_H^0(n)=B/n^2-J_H^0(n)$; its symbol satisfies
$Q_H^0(k)=\sum_nD_H^0(n)(1-\cos nk)\ge0$.

\begin{theorem}[Exact positive representation]
\label{s3:thm:positive-core}
Fix $a\in C_c^\infty((1/4,1/3))$, nonzero and nonnegative. There are
constants $\Lambda_{\rm c},\epsilon_*,H_*$ depending only on $a,\phi$ such
that, for $H\ge H_*$ with $\epsilon_H=E(\lceil H/16\rceil)\le\epsilon_*$,
\begin{equation}
 J_H(n)=J_H^0(n)+(J_n-B/n^2)\ind_{n<\lceil H/16\rceil}
       -\Lambda_{\rm c}\epsilon_HBH^{-2}a(n/H)
 \label{s3:positive-core-definition}
\end{equation}
is nonnegative, has range $H$, and, with $D_H=J-J_H$, satisfies
\begin{equation}
\begin{gathered}
|D_H(n)|\le CB/H^2\quad(n\le H),\qquad \|D_H\|_1\le CB/H,\\
 \|D_H-D_H^0\|_1\le CB\epsilon_H/H,\qquad
 Q_H(k):=\sum_nD_H(n)(1-\cos nk)\ge0.
\end{gathered}
 \label{s3:positive-core-bounds}
\end{equation}
The covariance $\widehat C^D(k)=2Q_H(k)/(B\lambda(k))$, understood on neutral charges,
$\lambda(k)=4\sin^2(k/2)$, gives, with
$\xi_i=\varphi_i-\varphi_{i-1}$ and $s_K=\sum K_n$,
\begin{equation}
 Z^f_{J,N}(h)=e^{\beta N(s_J-s_{J_H})}
 \E_{C^D}Z^f_{J_H,N}(h_i-i\sqrt{4\beta B}\,\xi_i/2).
 \label{s3:positive-core-source-identity}
\end{equation}
The identity holds for every finite source pattern and its derivatives,
with free endpoints; the decomposition is temperature independent.
\end{theorem}
\begin{proof}
The identity $\sum n^{-2}(1-\cos nk)=\pi|k|/2-k^2/4$ and
$\sum n^2J_H^0(n)\le CBH$ give $Q_H^0(k)\ge cB|k|$ for
$H|k|\le t_0$, with fixed small $t_0$. The uncompensated tail error is
at most $CB\epsilon_H\min(|k|,H^{-1})$. The margin has symbol at least
$c_aB/H$ when $t_0\le H|k|\le\pi H$: on bounded $Hk\ne0$ this follows
from $\int a(x)(1-\cos(Hk)x)dx>0$, uniformly on compact sets; at larger
$H|k|$, summation by parts bounds
$|H^{-1}\sum a(n/H)e^{ikn}|\le C/[H|\sin(k/2)|]$.
Choose $\Lambda_{\rm c}$ first and then $\epsilon_*$ small. This proves
$Q_H\ge0$. On the short head $J_H=J-D_H^0\ge J$; on the margin its
positivity follows from $a_0-\Lambda_{\rm c}\epsilon_*\|a\|_\infty>0$.
All other claims in \eqref{s3:positive-core-bounds} follow by summation.
The gradient covariance has diagonal $2\sum D_H/B$ and off-diagonal
$-D_H(n)/B$. Its characteristic function is
$\exp[-\beta N(s_J-s_{J_H})+\beta\sum_{a<b}D_H(b-a)\sigma_a\sigma_b]$.
Multiplication by the core weight proves
\eqref{s3:positive-core-source-identity} termwise, locally uniformly with
every fixed-order source and temperature derivative.
\end{proof}

\subsection{Critical core preparation}
\label{s3:core-thermodynamics}\label{sm:ent:records}

We use Ref.~\cite{Jump}, Supplemental Material, Lemma~S2.1,
Theorem~S2.2, Proposition~S3.2 and Lemmas~S3.3--S3.4 for diffuse deletion,
primitive energy bounds and the amplitude ceiling. At $\beta=\beta_c$ put
$m=m_c$, $\Delta=\beta_cBm^2=1/2$ and
$F_J(N)=NP_J-\log Z^f_{J,N}(0)$.

\begin{proposition}[Critical core plateau]
\label{s3:prop:core-plateau}
There are dyadic $H_k=2^{r_k}$ with $r_k/k\to1$ such that
$F_J(N)=\Delta\log N+o(\log N)$ and
\begin{equation}
 H_k(P_{J_{H_k}}-P_J)\longrightarrow0.
 \label{s3:core-pressure-plateau}
\end{equation}
On every fixed multiple of $H_k$, the physical free laws and the free
and stationary symmetric laws of
$J_{H_k,t}=J+t(J_{H_k}-J)$ have primitive limits with fair constant
slopes $\pm m$, uniformly for $0\le t\le1$.
\end{proposition}
\begin{proof}
GKS~\cite{GKS}, cross-bond interpolation and plus-state clustering give
$0\le F_J(N)\le\beta_c\sum_n\min(n,N)J_n\langle\sigma_0\sigma_n\rangle^+$,
hence upper slope $\Delta$. Delete
$D_r(n)=(1-\eta)B\ind_{n>R_\eta}\int_0^rue^{-un}du\le J_n$.
Its mass is $(1-\eta)Br+O_{R_\eta}(r^2)$ and escapes finite sets, so its
pressure loss is $(1-\eta)\Delta r+o(r)$. At $r=V/N$, the cited energy
comparison gives, for a primitive slope $mz$, $|z|\le1$, along backward
records with increments at most $s(j+C)\log2+O(1)$,
\begin{equation}
\begin{gathered}
 \int_0^V\mathcal D(v)dv\le2^{-j}V+(s/\Delta)(j+C)\log2+C',\\
 \mathcal D(v)=1-\E\langle z,P_vz\rangle\ge(1-e^{-v})/v,
 \qquad P_v=(v/2)e^{-v|\cdot|}*.
\end{gathered}
 \label{sm:ent:budget}
\end{equation}
Here $z$ is zero extended. First take volume at fixed $V,j,\eta$;
the omitted head costs $O(R_\eta V^2/N)$ and lattice errors
$O((V^2+V^3)/N)$. Then let $\eta\downarrow0$; the remaining constants
are uniform for $\eta\le1/2$. Running minima with $s<\Delta$ contradict
the lower bound at $j=\lceil\log_2V\rceil$, proving the slope.
Choose $r_k$ minimizing $F_J(2^j)-(\Delta+\varepsilon_k)j\log2$ for
$j\le k$, with $\varepsilon_k\downarrow0$ slowly enough that
$\max_{j\le k}|F_J(2^j)-\Delta j\log2|=o(k\varepsilon_k)$.
Then $r_k/k\to1$ and its backward increments are at most
$(\Delta+\varepsilon_k)j\log2$. GKS gives nonnegative decreasing increments of $F_J$; for $T\ge1$,
$F_J(TH)-F_J(H)\le2(T-1)[F_J(H)-F_J(H/2)]$, with integer rounding
understood. The backward budget therefore holds at each fixed $TH_k$, up to a
$T$-dependent constant. The cited interface bound gives $|z|=1$ and
$\E\operatorname{TV}_{\R}z\le2$. The external jumps cost two, forcing
constant slope on each fixed multiple.

On $TH_k$ sites, \eqref{s3:positive-core-bounds} bounds the
$J_{H_k,t}$ density ratio above and below uniformly, preserving the free
primitive limit. GKS bounds stationary second moments below by this value.
The pressure ceiling, $\|J_{H,t}-J\|_1=O(H^{-1})$, and $O(\log H)$ boundary
cost bound their amplitude above by $m$. Thus stationary limits also
saturate uniformly in $t$. The scaled canonical defect tends to
$k_*(x)=[x\phi'(x)-\phi(x)+1]/x^2$, whose integral is zero;
the extra $\ell^1$ error costs $O_T(\epsilon_H)$.
Consequently $H\sum_nD_H(n)\langle\sigma_0\sigma_n\rangle_{H,t}\to0$:
first average on $RH$ sites, with truncation error
$O((1+\log R)/R)$, then let $R\to\infty$. Pressure interpolation proves
\eqref{s3:core-pressure-plateau}.
\end{proof}

\subsection{Heat shells and the complete forward tail}
\label{s3:heat-tail-base}

For the full interaction, and for the prepared residual respectively,
\begin{equation}
 d_J(k)=2\sum_nj_n(1-\cos nk),\quad \widehat C_J=d_J/\lambda,
 \qquad \widehat C^D=\frac{2\sum_n(D_H(n)/B)(1-\cos nk)}\lambda.
 \label{s3g:full-symbol}
\end{equation}
Choose even $\chi\in C_c^\infty((-\pi/2,\pi/2))$, equal to one near zero,
$0\le\chi\le1$, and $P_R(k)=\chi(k)^2e^{-R^2\lambda(k)}$. The exact split is
\begin{equation}
 C^D=(1-P_R)C^D+P_RC^D,\qquad C_R^{\rm sh}=(P_R-P_{LR})C^D.
 \label{s3g:heat-split}
\end{equation}
Both terms are positive; write $C_R^{\rm ret}=P_RC^D$.
After $x\mapsto Rx$, the canonical shell and retained densities are $\pi(e^{-\xi^2}-e^{-L^2\xi^2})/|\xi|$ and
$\pi e^{-\xi^2}/|\xi|$; the shell variance is $b=\log L$.
For $d_n=D_H(n)/B$ and $\epsilon_n^D=n^2d_n-1$, put
\begin{equation}
 \sigma_R^D=R^{-1}+R^{-1}\sum_{n<R}|\epsilon_n^D|
                +\sup_{n\ge R}|\epsilon_n^D|.
 \label{s3:prepared-covariance-distance}
\end{equation}
The same expression with $\epsilon_n$ is $\sigma_R^G$; for $R\ge H$,
$\sigma_R^D\le\sigma_R^G+A_H/R$, where $A_H=B^{-1}\sum_{n\le H}n^2J_H(n)$.

\paragraph{Uniform kernel estimates.}
\label{s3:prop:prepared-covariance-distance}
Put $w(x)=(1+x^2)^{-1}$ and
$\|f\|_{m,w}=\sup_x|\partial^mf(x)|/w(x)$. For fixed $m\ge0$,
in rescaled coordinates,
\[
 \|C_R^{\rm sh}-C_*^{\rm sh}\|_{m,w}
 +\|C_R^{\rm ret}-C_*^{\rm ret}\|_{m+2,w}\le C_{m,L}\sigma_R^D,
 \qquad \|(C_R^{\rm ret}-C_*^{\rm ret})'\|_\infty\le C_L\sigma_R^D.
\]
On unit windows $I,J$ with centers $x_I,x_J$, the localized $H^4$ nuclear
bound is $C_Lw(x_I-x_J)$, multiplied by $\sigma_R^D$ for a difference.
On occupied $X$, operator and trace bounds are $C_L$ and $C_L|X|$,
with the same difference factor.

With
$\lambda_R(\xi)=R^2\lambda(\xi/R)$ the smooth multipliers
$\chi(\xi/R)^2(e^{-\lambda_R}-e^{-L^2\lambda_R})/\lambda_R$ and
$\chi(\xi/R)^2e^{-\lambda_R}\xi^2/\lambda_R$ have uniformly Schwartz
inverse transforms $f_R,g_R$, converging with error $O(R^{-2})$.
The shell and retained second derivative are respectively
$\mathcal L_R^df_R$ and $-\mathcal L_R^dg_R$, where $\mathcal L_R^af=R\sum_na_n[2f-f(\cdot+n/R)-f(\cdot-n/R)]$.
In each weighted derivative norm, Taylor's formula bounds $n<R$ by
$CR^{-1}\sum n^2|a_n|$. For $n\ge R$, use
$(1+x^2)/t^2\le C[1+(x-t)^2]$, $t\ge1$, and Schwartz summation to obtain
$C\sup n^2|a_n|$. The canonical Riemann-sum error is $O(R^{-1})$ by the
integrable $t$ derivative of $[2f-f(\cdot+t)-f(\cdot-t)]/t^2$. The first retained
derivative follows from
$|d_J(k)-d_0(k)|\le k^2\sum_{n<N}|\epsilon_n|+CE(N)|k|$.
The current jet kernel is
$(-1)^{s+1}\partial^{r+s+2}C$, $0\le r,s\le4$.
Schur summation and diagonal derivatives give the operator and trace
bounds. Localized Fourier coefficients are bounded by
$Cw(x_I-x_J)/[(1+r^2)(1+t^2)]$; their absolutely summable rank-one
expansion gives the nuclear bound. No scalar-field anchor is used.

\paragraph{The driven head memory.}
\label{s3:prop:heat-base-memory}
The exact temperature-independent base is
$\omega_R=(\nu_R,e_R,\delta_R)$, where
$\nu_R=R\sum_{n<R}\epsilon_n^D n^{-2}\delta_{n/R}$,
$e_R(t)=\epsilon^D_{\lfloor Rt\rfloor}$ for $t\ge1$, and $\delta_R=R^{-1}$.
Using $\|\nu\|_2=\int t^2d|\nu|(t)$, its shift is
\begin{equation}
 \nu_{LR}=L(s_{1/L})_*\left(\nu_R+
 R\sum_{R\le n<LR}\epsilon_n^Dn^{-2}\delta_{n/R}\right),\quad
 e_{LR}(t)=e_R(Lt),\quad\delta_{LR}=\delta_R/L,
 \label{s3:heat-base-shift}
\end{equation}
so
\begin{equation}
 \|\nu_{LR}\|_2\le L^{-1}\|\nu_R\|_2+
 (1-L^{-1})\sup_{n\ge R}|\epsilon_n^D|.
 \label{s3:heat-head-memory}
\end{equation}
The contracting head and vanishing future tail give $\sigma_R^D\to0$;
the base retains both, not just the variance error.

\paragraph{A wall-energy comparison.}
\label{s3:prop:heat-triangle-energy}
For a spin path $S$ with $N_w$ walls, heat and positive triangular
retained energies differ by at most $C_pN_w$ at all sufficiently late scales.
Take the real Schwartz kernel
\[
 \widehat g_R(\xi)=\chi(\xi/R)e^{-\lambda_R(\xi)/2}
       \frac{\xi}{2R\sin(\xi/(2R))},\qquad u=g_R*S.
\]
Uniformly, $\int g_R=1$ and $\|g_R\|_1+\||x|g_R\|_1+\|g_R'\|_1\le C$.
Thus $\|S-u\|_1\le CN_w$ and $\|u'\|_2^2\le CN_w$.
The two energies are
$(pR/8)\sum_{n\ge1}j_n\|T_{n/R}u-u\|_2^2$ and
$(pR/8)\sum_{n\ge R}j_n\|T_{n/R}S-S\|_2^2$.
Short translations cost $C(n/R)^2N_w$; each difference of the squared
norms costs $CN_w$. Their difference is therefore bounded by
$CpN_w[R^{-1}\sum_{n<R}n^2j_n+R\sum_{n\ge R}j_n]$.
A prepared finite core adds $CA_HN_w/R$. For opposite endpoint spins
one subtracts the integrands before summing; the individual infrared
energies need not be finite.

\subsection{The physical quadratic normalization}
\label{s3:heat-quadratic-benchmark}

Let $C_R=(1-P_R)C_J$, $S_R=C_R(0)$, $V_R(n)=S_R-C_R(n)$, and
\begin{equation}
 I_R(p)=\sum_{1\le n<R}n[e^{-pV_R(n)}-e^{-pS_R}],\quad
 Y_R=Re^{-pS_R/2},\quad Q_R(p)=\frac{I_{LR}(p)-I_R(p)}{Y_R^2}.
 \label{s3:heat-pair-data}
\end{equation}
For the formal wall-counting parameter $z$, the complete current and
activity jets are
\begin{equation}
 [z^2]Z_R=\tfrac13e^{-pS_R}-2I_R(p),\qquad [z]y_R=Y_R,\qquad p_R=pZ_R^2.
 \label{s3:heat-complete-jet}
\end{equation}
The contact $1/3$ follows from
$\partial_\xi\log[e^{\xi S/2}(I+t\sigma_x)e^{\xi S/2}]|_0
=(\operatorname{arctanh}t/t)S$, $t=ze^{-pS_R/2}$, $S=\sigma_z$;
a distinct opposite pair has current jet $-2n$. Summing all pairs below $R$ includes the accumulated neutral feedback.

\begin{proposition}[Critical heat coefficient]
\label{s3:thm:heat-coefficient}
At $p=2$,
\begin{equation}
 Q_R(2)\longrightarrow b e^{-\gamma_E},\qquad
 \gamma_*:=4b^2e^{-\gamma_E}>0,
 \label{s3:heat-Q-value}
\end{equation}
where $-\gamma_*$ is the complete quadratic coefficient of
$q=b(p/2-1)$. Hence the activity normalization is
\begin{equation}
 v=2be^{-\gamma_E/2}y\quad\hbox{to first order}.
 \label{s3:heat-normalized-activity}
\end{equation}
\end{proposition}
\begin{proof}
Put $A_h=\gamma_E/2$,
$V(x)=\int_0^\infty(1-\cos\xi x)e^{-\xi^2}d\xi/\xi$ and
$K(x)=V(x)-\log x-A_h$. The inverse Fourier kernel $\check P_R$ of $P_R$ has sum one and
$|\check P_R(s)|\le C_NR^{-1}(1+|s|/R)^{-N}$.
For $T_l(n)=(l-|n|)_+$, $T_l-\check P_R*T_l$ is bounded by $Cl^2/R$ if
$l<n$, $Cl$ if $n\le l\le R$, and $CR$ if $l\ge R$.
Thus its $l^{-2}$ sum is at most $C_L[1+\log^+(R/n)]$.
The canonical symbol gives
$S_R^0=\log R+\gamma_E+1-A_h+O(R^{-1})$ and
$C_R^0(n)=K(n/R)+O(R^{-1}+n^{-2})$.
Splitting the defect at $N_0=\lfloor\sqrt R\rfloor$ proves, for
$N_0\le n\le LR$,
\[
 |C_R(n)-K(n/R)|\le C_{J,L}R^{-1/2}
       +C_LE(N_0)[1+\log^+(R/n)].
\]
For fixed small $\varepsilon>0$ and all $n<R$,
$C_R(n)\le C_{J,\varepsilon}+(1+\varepsilon)\log(R/n)$.
Write $c_R=S_{LR}-S_R$ and $U_R=c_R-C_{LR}+C_R$.
The heat bounds give $c_R\to b$ and
$U_R(n)=V(n/R)-V(n/(LR))+O(\sigma_R^G(n/R)^2)$.
Subtract the sums defining $Q_R$ before estimating: on $n<R$ the
bracket multiplying $n/R^2$ is
$e^{2C_R(n)}(e^{-2U_R(n)}-1)+1-e^{-2c_R}$.
Its quadratic cancellation makes the small-$n$ part vanish and gives
the integrable majorant $C(x^{1-2\varepsilon}+x)$; away from zero the
preceding profile estimate gives convergence of the Riemann sums.
To evaluate the limiting integral, cut both terms at $a>0$ and change
$x\mapsto x/L$ in the first. Their difference is
$\int_{a/L}^a x(e^{2K(x)}-1)dx\to e^{-2A_h}\log L$ because
$K(x)=-\log x-A_h+O(x^2)$.
Equation~\eqref{s3:heat-complete-jet} converts this to $-4b^2e^{-\gamma_E}$;
the contact difference divided by $Y_R^2$ is $O(R^{-2})$.
\end{proof}

\subsection{The fixed field and graph space}

Put $S=\sigma_z$, $\alpha=\sqrt p$, and
$\vartheta=h/\alpha-i\psi/2$, where $\psi=\partial_x\varphi$.
On a unit window $I$ use $E_I=H^4(I;\C)$ and the projective Taylor norm
\begin{equation}
 F(z)=\sum_{n\ge0}\langle f_n,z^{\otimes n}\rangle,
 \qquad \|F\|_{\mathcal F_\rho}=\sum_{n\ge0}\rho^n\sqrt{n!}\,\|f_n\|_\pi.
 \label{s3n:fock-norm}
\end{equation}
For several windows use $\prod_I\sqrt{n_I!}$; $I$ is active when $n_I>0$.
Port representations agree in the quotient if they coincide on fields
with matching traces through order three. Use a fixed smooth cover of
bounded multiplicity and its extension operator.
Under $S_Lf(x)=L^{-1}f(x/L)$ each leg costs $C/L$; at most $CL$ fine
windows meet a coarse window. All bounds are uniform for $0<\delta\le1$.

\begin{lemma}[Field estimates]
\label{s3n:lem:fock-embedding}
On fixed nested radii $\rho>\rho'>0$, translation and differentiation satisfy
\begin{equation}
 \|F(\cdot+h)\|_{\rho'}\le
 e^{\|h\|^2/[2(\rho^2-\rho'^2)]}\|F\|_\rho,
 \qquad \|D^dF\|_{\rho';(H^{-4})^{\otimes_\pi d}}
 \le C^d\sqrt{d!}\,\|F\|_\rho.
 \label{s3n:projective-ports}
\end{equation}
Positive or signed Gaussian operators with uniformly bounded operator
norm and trace norm $C|Y|$ cost $e^{C|Y|}$, where $Y$ comprises only
active windows. Their inverses are bounded on a further fixed reserve.
An entire $H^2$ envelope $|F(z)|\le M e^{a\|z\|_{H^2}^2}$ embeds into
\eqref{s3n:fock-norm} for sufficiently small $a$ at fixed $\rho$.
\end{lemma}
\begin{proof}
Binomial expansion and Cauchy--Schwarz, with $t=\rho'/\rho$, give
translation; $\sum_m\binom{m+d}{d}t^{2m}=(1-t^2)^{-d-1}$ gives the port bound.
Also $|F(z)|\le\|F\|_\rho e^{\|z\|^2/(2\rho^2)}$.
Gaussian integration therefore costs the determinant
$\det(1-Q/(\rho^2-\rho'^2))^{-1/2}\le e^{C|Y|}$ when the squared-radius
gap exceeds $\|Q\|_{\rm op}$ by a fixed factor. For signed $Q=Q_+-Q_-$,
use independent real and imaginary Gaussians and two reserves. The
opposite signs compose to the identity by the mean-value property of a
circular complex Gaussian; finite-rank approximation and the determinant
bound justify the limit. Degree-zero windows are identities throughout.
Finally $H^4\hookrightarrow H^2$ is nuclear, with singular values
$O((1+k^2)^{-1})$ after a fixed reflected extension. If its nuclear norm
is $N_i$, Cauchy and polarization give coefficient majorant
$CMn^{1/4}(e\rho N_i\sqrt{2a})^n$; its sum converges for
$e\rho N_i\sqrt{2a}<1$.
\end{proof}

All ordinary and marked coordinates remain independent until reduction.
The shell and retained current covariances obey the exact cocycle
$Q_j=\Gamma_j+S_LQ_{j+1}S_L^*$. Wick charts are actual signed Gaussian
operators from the lemma. Their generating formula on one charged rod is
\[
 \mathcal W_j^qe^{\langle\vartheta,r\rangle}
 =e^{i\alpha q\varphi(a)}e^{\langle\vartheta,r\rangle}
   e^{\langle r,Q_jr\rangle/8-\alpha q\langle\partial C_j(\cdot-a),r\rangle/2}.
\]
The cocycle gives current dilation and multiplier $e^{-pC_j^{\rm sh}(0)/2}$.
For charge variance $Q=q^TC^{\rm sh}q$, positivity gives the mixed shift
$\|h_q\|^2\le |p|CQ/4$. A fixed radius gap makes translation times
$e^{-(\Re p)Q/2}$ at most one, uniformly in the number of bare points:
no determinant is charged per wall.

For ordinary rods $q_a\in\{-1,0,1\}$, put $N_q=\sum_aq_a^2$;
half charges remain in the endpoint modules. For separated rods,
$K(r)=V(r)-\log|r|-\gamma_E/2$ obeys
$|K^{(m)}(r)|\le C_m(1+|r|)^{-m-2}$ at $|r|\ge1$ and
$C_*^{\rm sh}(r)=K(r/L)-K(r)$. The current kernel is $V''$ on a rod
and $K''$ between rods; the mixed kernel is $-V'(x-a)$ on the anchor's
rod and $-K'(x-a)$ otherwise. Multiply the local chart by
$e^{-p\sum_{a<b}q_aq_bK(a-b)}$. Schur and nuclear estimates give
operator bound $C$, active trace $C|Y|$, and mixed squared norm
$C\sum q_a^2$.

For compatible old rods with fine-coordinate gaps at least one, the
same bounds hold for $(x-z)^{-2}$ and $\sum_aq_a/(x-a)$ between rods;
the mixed sum excludes the anchor's own rod. Derivatives have summable rows.
For the undifferentiated mixed kernel, replace anchors by disjoint unit
bumps and use the Hilbert-transform multiplier $-i\pi\operatorname{sgn}\xi$;
the point-minus-bump error is $O(|x-a|^{-2})$. Thus its $H^4$ squared norm
is $C\sum q_a^2$, without $\log L$. The field lemma bounds both charts
and inverses by $e^{C(|Y|+N_q)}$. Differentiating the covariance cocycle
on exponential generators gives
\begin{equation}
 \E_*\mathcal W_XF=e^{-pN_qb/2}\mathcal W_{X/L}(F\circ S_L)
 \label{s3n:cross-wick-intertwining}
\end{equation}
for rods remaining separated after dilation, hence on the completion.
Inactive windows cost nothing. General tails enter only through
shell-operator differences, without an additive potential constant.

\begin{definition}[Pointwise rod norm]
\label{s3n:def:tree-norm}
Each occupied rod has length at least one, a matrix Fock coefficient,
attachment windows and a net-charge phase. Bare wall points are separate
inputs. Each edge carries charge or current ports and has scalar modulus
or nuclear norm at most $a_ew(x_{e_+}-x_{e_-})$.
For graph representations with pointwise majorant
$a_G\prod_ew(x_{e_+}-x_{e_-})$, set
\begin{equation}
 \|R\|_{A,B_{\rm r},\rho}=\inf_{\rm representations}\sup_{\rm root}
       \sum_G e^{A|X_G|+B_{\rm r}|V(G)|}a_G.
 \label{s3n:tree-norm}
\end{equation}
The infimum fixes the labelled projective tensor. The majorant retains
relative positions and $a_G$ the expansion factorials. $V(G)$ counts unfused constituents, including those in closed
blobs, and excludes Wick edges. Choose $A,B_{\rm r}>0$ independently of
the tail constant $B$. A bridgeless block, possibly with parallel edges, may use its anchored
second shape moment, the integral of the coefficient norm over relative
positions with weight $1+\operatorname{diam}(X)^2$, keeping its
rods, attachments and constituent count.

Store actual kernel and contracted primitive separately on inherited
cuts. For charge $C$ and current $D$, retain charge-point values on
$CC$, one primitive and its oriented difference on $CD$, both $k$ and $A=-k''$ on $DD$ in the $w$ norm (second
difference on the lattice). All cuts keep their difference flags,
contact values and reflection grading. Use continuous representatives in the scalar or
nuclear weighted supremum norm, and fine values on the lattice.

Complete the algebraic space of labelled projective tensors modulo the
null space of
\eqref{s3n:tree-norm}. Cut derivative and reflection identities impose
closed linear constraints. Finite-volume chronological evaluation
descends continuously. Equal evaluations alone do not identify
distinct derivative or primitive data.
Marks are microscopic roots with source Taylor factorials. The ordinary
even and odd sectors are $\mathcal B_0,\mathcal B_1$. Endpoint leading vectors are
finite-dimensional and unrestricted in size.
\end{definition}

Sum successive unpinned vertices with $\delta\sum_{\delta\Z}w\le C$
and use $w*w\le Cw$ along paths. The resulting $C^{|V(G)|}$ is absorbed
in the constituent weight. Thus the norm dominates anchored incidence
and the complete-source envelope used in reconstruction.

\subsection{The common map}

For iteration put $\omega_j=\omega_{H_j}$ and $R=(R_0,R_1)$,
$\|R\|=\|R_0\|+\|R_1\|$.

\begin{theorem}[Exact common microscopic maps]
\label{s3n:thm:common-map}
Fix a complex current disk $|p-2|<1/4$ with reserved smaller disks.
There exist one integer $L$, fixed field and parameter reserves, weights
$A,B_{\rm r}$, and radii $r_*,\sigma_*>0$ such that every prepared positive heat
base with $\sigma_j^{\rm op}:=\sigma_{H_j}^D<\sigma_*$ admits maps
$F[\omega_j]:(p,y,R_0,R_1)\mapsto(p',y',R_0',R_1')$ on
$|y|+\|R_0\|+\|R_1\|<r_*$. They preserve the finite-volume partition, free endpoints, all finite
microscopic source patterns and source-dependent scalars. The complete
source-independent scalar, including its boundary part, is external.
They are holomorphic with uniform current derivatives through order six
on smaller fixed charts; $\partial_\beta\omega_j=0$.

The zero-wall sheet has no drift. The exact singleton multiplier is
\begin{equation}
 \lambda_j(p)=L e^{-pC_j^{\rm sh}(0)/2}.
 \label{s3n:common-singleton}
\end{equation}
Write $a_j=C_j^{\rm sh}(0)-b$; the heat estimates give $a_j\to0$.
Charge parity is preserved, and the charged proper output has no linear
term in $y$. Set $A_{\mu,j}=D_{R_\mu}R_\mu'(p,0,0,0)$, $\mu=0,1$, and $A_j=A_{0,j}\oplus A_{1,j}$. Then
\begin{align}
 \|A_{\mu,j}\|&\le\kappa<1/4,
 \label{s3n:ordinary-gap}\\
 \|R'\|&\le\kappa\|R\|+C_L(|y|^2+|y|\|R\|+\|R\|^2),\quad
 \|D_RR'-A_j\|\le C_L(|y|+\|R\|).
 \label{s3n:ordinary-map-bounds}
\end{align}
More precisely, the charged nonlinear birth is bounded by
$C_L[|y|^3+|y|\|R_0\|+(|y|+\|R\|)\|R_1\|]$.
Inherited charged inputs retain their linear activity contribution.
For a rooted module $E=E_{\rm lead}+E_{\rm prop}$, one- and two-root
proper blocks have the same gap without bulk root placement, with forcing bounded by
$C_Lr(\|E_{\rm lead}\|+\|E_{\rm prop}\|)$,
$r=|y|+\|R\|$; two-root maps retain products of one-root modules and
coincident insertion contacts.

Put $q=b(p/2-1)$ and define, on the zero-wall sheet,
\begin{equation}
 B_j=\tfrac12D_y^2R_0',\qquad
 d_j=\tfrac12D_y^2q',\qquad \Lambda_j=D_{R_0}q'.
 \label{s3n:actual-quadratic-data}
\end{equation}
Then
\begin{equation}
 \|A_{0,j}-A_{0,*}\|+\|A_{1,j}-A_{1,*}\|+
 \|B_j-B_*\|+|d_j-d_*|+\|\Lambda_j-\Lambda_*\|
 \le C_L\sigma_j^{\rm op},
 \label{s3n:microscopic-jet-distance}
\end{equation}
also after three current derivatives. The comparison concerns these jets; it asserts no full nonlinear
operator-norm convergence on inherited primitives. Current extraction is the exact physical rebasement
\begin{equation}
 Z=1+\Delta\alpha/\alpha,\qquad p'=pZ^2,\qquad h'=LZh,
 \qquad\vartheta_{\rm old}(x)=L^{-1}\vartheta_{\rm new}(x/L).
 \label{s3n:physical-current-rebase}
\end{equation}
In particular $p_j=4\beta B(Z_j^{\rm phys})^2$ with the original source.
\end{theorem}

\begin{proof}
We work on finite chronological tensors and extend by the following
absolute bounds.

\paragraph{Exact connected integration.}
\label{s3n:thm:finite-shell}\label{s3n:prop:projective-forest}
Form Gaussian cumulants in matrix tensor space before chronological
evaluation, also for interleaved rods. The connected output is
$\sum_{m\ge1}(m!)^{-1}\sum_{X_1,\ldots,X_m\ {\rm disjoint}}
\kappa_\Gamma(K_{X_1},\ldots,K_{X_m})$.
Apply the Banach-valued forest formula and positivity theorem
\cite[Theorems~III.1 and IV.5]{Forest} to finite-rank positive covariances,
then pass to the limit using the field lemma. A degree-$d$ ordinary
vertex costs $C^dd!$ and each covariance edge $C_Lw$.
A bare site is kept exactly:
$e^{y\delta\sigma_x}=\cosh(y\delta)[I+\tanh(y\delta)\sigma_x]$.
Its point ports have exponential rather than factorial branching and
$\delta\sum_aw(a-z)\le C$; no regulator is charged per bare point.
\label{s3n:lem:rooted-tree-sum}
The ordinary and bare child sums are $(1-Ct)^{-1}$ and $e^{Ct}$.
Their rooted depth truncations satisfy $z_{n+1}\le\Phi(z_n)$, where
$\Phi(z)=C_Lr(1-C_Lz)^{-2}e^{C_Lz}$ and $\Phi(2C_Lr)\le2C_Lr$ for small $r$.
Thus the series converges absolutely,
with norm $C_Lr$, nonlinear remainder $C_Lr^2$, and marked
covariance-difference bound $C_L\sigma_j^{\rm op}r$. One-point terms resum to
$\delta'^{-1}\operatorname{arctanh}
[e^{-p\nu/2}\tanh(\delta y)]$, $\nu=C_j^{\rm sh}(0)$, $\delta'=\delta/L$,
with multiplier \eqref{s3n:common-singleton} and odd remainder
$O_L(\delta^2y^3)$. Retain all $\cosh$ scalars. Thicken only connected
outputs containing two bare points or an ordinary input; resum
independent gap points once.

\paragraph{Spatial closure and the two primitive norms.}
\label{s3:cor:bridge-dressing}
The Fourier transform
$\widehat w(\xi)=\int e^{-i\xi r}w(r)dr=\pi e^{-|\xi|}$ gives, for $n\ge1$,
\[
 w^{*n}(r)=\frac{n\pi^{n-1}}{n^2+r^2},\qquad
 \int r^2w(r)w^{*n}(r)dr=\frac{n\pi^n}{n+1}.
\]
These identities give $w*w\le Cw$ and the cycle second-moment bound.
For a bridgeless $m$-vertex graph, fix a spanning tree. Each of its
edges belongs to a cycle, so retain a spanning unicyclic subgraph
whose cycle contains that edge and bound all other edge weights by
one. Sum its attached trees in $L^1$ and convolve the complementary
cycle path; the displayed integral bounds the squared length of the
chosen edge. The tree-path inequality bounds the squared anchor diameter by
$(m-1)$ times the sum of squared edges of the fixed spanning tree,
giving anchored second moment $C^mm^2$. Attachment offsets and extents
within the retained rods contribute polynomial length factors, absorbed
once by the occupied-support reserve.

The bounds are uniform on meshes $0<\delta\le1$ with measure
$\delta\sum$: replace each lattice point by its length-$\delta$ cell.
The resulting shifts of edge arguments by at most one change $w$ by
a fixed factor, and the diameter weights are comparable. Make this
comparison only after retaining the unicyclic subgraph, so its cost
is $C^m$, independent of the discarded extra edges. The same argument
works with one edge $(1+|r|)^{-s}$, $s>1$: convolve along the
complementary path and use
$\int r^2w(r)(1+|r|)^{-\min(s,2)}dr<\infty$.
For any coefficient with actual second moment $M_2(a)=\int(1+s^2)|a|(ds)$,
\begin{equation}
 (|a|*w)(r)\le2M_2(a)w(r).
 \label{s3n:bridge-dressing}
\end{equation}
Only actual attachment offsets enter this estimate.

Interpolate the cross covariance of two independently integrated
unmarked neutral clusters $A,B$ by $t$. In
$G(1)-G(0)=G'(0)+\int_0^1(1-t)G''(t)dt$,
neutrality and spin inversion make the zero-current components of their
complete one-current responses $\lambda_A S,\lambda_B S$.
Other one-link terms retain at least one current degree in the ordinary
forest bound: \eqref{s3n:tree-scale} gains at least $L^{-1}$ for two
separated rods; close fusions first extract the local scalar and uniform
current jets. Contract the zero-current dual coefficients now, retaining
both
\[
 k_{AB}(r)=-\tfrac14\langle\lambda_A(x),C(r+x-z)\lambda_B(z)\rangle,
 \qquad A_{AB}=-k_{AB}'',
\]
where $C$ is the fresh shell covariance. The two cross links in the
remainder close a cycle through the connecting paths inside $A$ and
$B$. Sum the attached trees in $L^1$. The cycle bound for the remaining
bridgeless block controls its attachment offsets without a second moment
for either whole side of the cut. Later operations act within the two
sides and never reopen a dual current port on inherited $k$.

For fresh kernels choose real even $\zeta\in C_c^\infty((-\pi,\pi))$,
equal to one on their Fourier band. At $t=\delta\xi$, the multipliers
\[
 m_+(t)=\frac{\zeta(t)it}{e^{it}-1},\qquad
 m_-(t)=\frac{\zeta(t)it}{1-e^{-it}},\qquad
 m_{DD}(t)=m_+(t)m_-(t)=\frac{\zeta(t)^2t^2}{4\sin^2(t/2)}
\]
intertwine continuous derivatives with the respective lattice differences.
Their Schwartz kernels preserve the $w$ norm and commute with dilation.
Use the $O(\delta^2k'')$ approximation only for fresh smooth kernels.
On an inherited primitive use
$\chi k''=(\chi k)''-2\chi'k'-\chi''k$, where $k'$ on the fixed collar follows by Taylor's formula from the two
stored norms. Discretely,
$\chi\Delta_\delta k=\Delta_\delta(\chi k)
-D_+[(D_-\chi)k]-D_-[(D_+\chi)k]+k\Delta_\delta\chi$.
Thus only single divergences and compact contacts are added. Under
dilation $A(r)\mapsto L^2A(Lr)$ and $k(r)\mapsto k(Lr)$; separated primitives gain
$L^{-2}$, but the contact value $k(0)$ is retained.

\paragraph{The invariant grading of every old cut.}
\label{s3n:cut-invariance}
Reflection reverses rod order and attachments and transposes matrices.
Since $\varphi$ is a pseudoscalar, $\psi,\vartheta$ are scalars. Spin
inversion acts by $K[\vartheta]\mapsto\sigma_xK[-\vartheta]\sigma_x$.
The reflection/inversion signs are $(-,-)$ for $[S,\cdot]/2$, $(+,-)$
for a current derivative and $(-,-)$ for its spatial derivative.
The $n_C,n_D$ open ports have signs
$((-1)^{n_C},(-1)^{n_C+n_D})$. The edge derivatives contribute
$(-1)^{n_D}$ under reflection, making every closed pair even. Fresh contractions, signed charts, reflected chronological
words and subsequent bare insertions preserve this grading. In
$c=1+(c-1)$ the unit preserves its bridge and the compact cross link
closes its block. Average each extraction with its reflection, keeping
both fine anchors.

The boundary port splits into charge and dipole,
\[
 \delta_bSK-\delta_aKS=
 \tfrac12(\delta_a+\delta_b)[S,K]+
 \tfrac12(\delta_b-\delta_a)\{S,K\}.
\]
Two-port self contacts are even and the full commutator obeys Leibniz.
An unmarked neutral zero-current spin leaf has $S=S^T$. The grading
therefore makes its charge measure $\mu$ odd, of mass zero. On its occupied rod
$I$, the cumulative charge $\nu(t)=\mu((-\infty,t])$ satisfies
$\int\|\nu\|\le |I|\|\mu\|_{\rm TV,\pi}$. Both cumulative charges enter the $CC$ bridge,
\[
 k_{\rm new}(r)=\iint k(r+s-t)\nu_A(s)\nu_B(t)\,ds\,dt,
 \qquad A_{\rm new}(r)=-k_{\rm new}''(r).
\]
Use one such integration for $CD$ and none for $DD$, which already has
both primitives. Lattice cumulative sums carry oriented one-cell
offsets. Only kernel values enter. Count each local length once. A long blob retains rods and second moment until bounded-shape
localization, with moment ratio $CL^{-2}$ beyond a fixed multiple of
$L$ in diameter.

Latent scalar branches sum since $\sum w<\infty$. A path stores
\[
 (A_{13},k_{13})=(A_{12}*a_2*A_{23},\ k_{12}*a_2*A_{23}).
\]
Differences commute with convolution. For flags at opposite ends,
convolve the primitives before taking both differences. No entire-side
moment enters. Lost weight pays for an absorbed
constituent, which otherwise remains latent. Pure compatibility bridges lack the odd port needed for spin-leaf
projection. Retain all source scalars and marked roots.

\paragraph{Exact Ward normalization, including contacts.}
\label{s3:lem:lattice-charge-ports}
In later-on-left convention the reference generator is
$\mathcal A=y\sigma_x+\xi S$. The chronological derivative of $S_x$ inserts
$[S,\mathcal A]=2iy\sigma_y$ and $\delta(x-x_k)[S,O_k]$ at collisions. The second self contact is
$[S,[S,\mathcal A]]=4y\sigma_x$. On the lattice,
$\mathcal QK=[S,K]/2$ satisfies $\mathcal Q^2K=(K-SKS)/2$,
$\mathcal Q^{2n+1}=\mathcal Q$ and $\mathcal Q^{2n}=\mathcal Q^2$ for $n\ge1$.
For $K=e^{-y\delta}e^{y\delta\sigma_x}e^{hS}$ the first two are
$i e^{-y\delta}\sinh(y\delta)\sigma_y e^{hS}$ and
$e^{-y\delta}\sinh(y\delta)\sigma_x e^{hS}$: both are single-wall
$O(y\delta)$ contacts. Telescoping
$U(N,i)S U(i,0)$ gives the exact commutator identity and the external
half charges; diagonal source and marked insertions commute with $S$.

A rod $[a,b]$ retains its diagonal holonomy
$D(b,c)K_cD(c,a)$, $D(b,a)=e^{S\int_a^b\xi}$, and boundary port
$\delta_bSK-\delta_aKS$. Two integrations by parts against $k$ give
bulk charged ports, rod contacts, external endpoints, the singleton
$4k(0)y\int\sigma_x$, and the same-rod contact
$2[k(0)K-k(b-a)SKS]$, together with cutoff derivatives. The neutral
boundary dipole is retained. A finite-width diagonal jet differentiates
as $-y\sigma_xDM+DM y\sigma_x$ between its two outside propagators;
the diagonal terms cancel because the holonomy was retained. This does
not translate an arbitrary Fock argument. Returning dipoles use only
actual values through
\[
 -\partial_r^2\int_0^{\ell_A}\!\int_0^{\ell_B}k(r-s+t)dsdt
 =k(r)-k(r+\ell_B)-k(r-\ell_A)+k(r-\ell_A+\ell_B).
\]

Let $u=\|k\|_{0,w}+\|A\|_{0,w}$ and $\varepsilon=r+u$. The full contact forest
has branch factor at most $\prod_{I\subset X}(1-C\varepsilon)^{-2}
\le e^{4C\varepsilon|X|}$, paid by one occupied-support reserve.
Repeated returns are bounded by $\sum_{n\ge0}(C_1\varepsilon)^nu\le2u$.
In \mbox{$Z(K+P)-Z(K)=\int_0^1DZ(K+tP)[P]dt$}, the $n$ choices cancel
$1/n$; different components join and same-component returns close cycles.
Retain linear same-rod terms and external endpoints. Passive scalars are removed by their
conditional connected logarithm
\[
 \log Z_{S|A}-\log Z_S
 =\sum_C\Phi_S(C)\left[\prod_{s\in C,a\in A}c(s,a)-1\right].
\]
It costs $C\|S\|$ and $C\|S\|\|A\|$ in the attached output. Retain the complete finite-volume scalar, including any boundary
logarithm, and all source-dependent scalar jets.

\paragraph{Fusion and the strict pointwise gain.}
\label{s3:lem:current-fusion}\label{s3n:lem:tree-scale}
In a closed alternating word of separated ordinary inputs, deleting a
closest adjacent pair at $d\ge1$ changes each side's logarithmic
interaction by at most $\log2$. Iteration gives $C^M\prod d^{-\Re p}$. In a true fusion group the coarse span is at most
$C(M+|X|/L)$. The bounded alternating primitive and $\|V''\|_1<\infty$
therefore bound the coarse smooth variogram factor by
$e^{C(M+|X|/L)}$. An odd word leaves its actual unpaired site, with
cross cost $O(\log(2+M+|X|/L))$, absorbed by the same bound.
Constituent-weight loss pays the $CM$ term; occupied-support shrinkage
pays $C|X|/L$. Separated groups use
\eqref{s3n:cross-wick-intertwining}. The net odd charge supplies $L^{-\Re p/2}$. The heat/triangle wall estimate above
bounds the odd-word variance below by
$\sum_{H\le n<LH}nJ_n/B-CM-CA_HM/H$, giving the same bound for the
actual late shell. Apply this estimate only after matrix multiplication closes an
alternating word; other words remain open tensors.

The signed fusion bound above costs $e^{C(|Y|+N_q)}$; current degree
pays the active-window term.

For local neutral proper coefficients, subtracting constant and linear
current jets supplies $M^2d_{\max}^2/L^2$. Pair matching and the one bulk
placement then give $C^M L^{-1+(2-\Re p)_+}$. For charged proper
coefficients the first subtraction supplies $Md_{\max}/L$;
$\prod d^{-\Re p}d_{\max}\le C^M$ because $\Re p>1$, giving
$C^ML^{-\Re p/2}$. For total current degree $D$ on $m$ active fine
windows, the projective multinomial bound is
$\sqrt{D!/\prod n_i!}\le m^{D/2}$, $m\le\min(D,CL)$.
Together with $(C/L)^D$, this gives $C^DL^{-D}$ for $D=2,3$, and
$C^DL^{-D/2}$ for $D\ge4$; it restores the field reserves after dilation.

For $m$ separated rods, $L^m$ placements and $m-1$ edges
$w(Lr)\le2L^{-2}w(r)$ give
\begin{equation}
 L^{2-m-N_q\Re p/2-D}.
 \label{s3n:tree-scale}
\end{equation}
If rods fuse to $m'$ groups and the old tree has $c$ internal components,
its $m-c$ internal links sum in $L^1$. A spanning selection of short gaps
has at most $2^{m-1}$ choices; the remaining $c$ anchors cost $L^c$ and
$c-1$ external links cost $L^{-2(c-1)}$. The result is
\begin{equation}
 C^mL^{2-c}\le C^mL^{2-m'}.
 \label{s3n:fusion-placement}
\end{equation}
The exceptions are scalar trees and the zero-current two-spin pair.
Remove scalars; the cut grading gives the pair a primitive or second
moment. Ward normalization has zero bulk linear output at $y=0$ and
retains endpoints. For a second-moment
blob, bounded-shape neutral localization gives $(D/L)^2$ before the
bulk factor $L$; its long part costs
$L\int_{D>L}|K|\le M_2(K)/L$. No $e^{yD}$ bound across a long hole is used. Omitting bulk root
placement gives the pinned proper gap; carry its leading vector exactly.

\paragraph{Local closure and the analytic selector.}
\label{s3n:extraction}
All anchors remain on $\delta\Z$ with measure $\delta\sum$ and
$\delta'=\delta/L$. Unit windows are norm covers. At each consecutive
ordinary gap $d$ insert $1=\chi(d/L)+(1-\chi(d/L))$, with smooth $0\le\chi\le1$, $\chi=1$
on $[0,1]$ and vanishes on $[2,\infty)$. Join the first-term gaps with
their complete chronological propagators. The remaining compatibility
is $c(d)=1-\chi(d)$ after dilation; because rods have minimum length one,
the adjacent product equals the all-pairs product. A group of $M$ rods
obeys $|X'|\le |X|/L+CM$. New connected joins cost $(C_Lr)^{M-1}$;
internal fusion is instead paid by lost constituent weight.

The unmarked local projectors are $cI$, $y\sigma_x$, and
$\Lambda(\vartheta)S$. Extract $\Lambda(1)$ using
$\mu_{\delta,a}=\delta^{-1}\int_a^{a+\delta}\vartheta$ and its reflected
average at the two actual rod endpoints. These dual norms are uniform,
and $\delta\sum_a\mu_{\delta,a}=\int\vartheta$. Retain the zero-mass jet
and all higher source coefficients. On a bounded shape,
$\Lambda(S_Lf)=L^{-1}\Lambda[f(\cdot/L)-f(a/L)]$ gains a further $L^{-1}$;
long shapes retain their moment. Marked tensors are not projected by
unmarked symmetry. Charged tie choices retain the actual unpaired site
and are averaged with their reflected-transposed partners.

Chronological Dyson regrouping rebases scalar, current and singleton
insertions into the reference transfer, including holes and endpoints.
For $t=(c,\Delta\alpha,\Delta y)$, remaining exclusions contain a
counterterm and occupied input, costing $C|t|r+C|t|^2$. Let $P$ project onto the scalar, current and singleton jets above. Solve
$\mathcal H(F,t)=PF-t+\mathcal E(F,t)=0$,
$\|D_t\mathcal E\|\le C(r+|t|)$.
Contraction on $|t|\le Cr$ gives the unique analytic selector on fixed
reserves, independently of hole lengths. Its current
counterterm is $(\alpha+\Delta\alpha)\vartheta S$, proving
\eqref{s3n:physical-current-rebase} on rods and endpoints simultaneously.

\paragraph{Choice of constants and actual-profile bounds.}
Choose field reserves first, using the uniform heat, signed-chart, and
fusion operator constants. Let $C_g$ bound the added windows per rod in
$|X'|\le |X|/L+C_gM$. Choose $A>0$ and $B_{\rm r}>2AC_g+C_{\rm fus}$,
with $C_{\rm fus}$ larger than twice the fixed per-rod fusion costs.
For large $L$, support shrinkage pays $e^{C|X|/L}$; a true $M$-rod
fusion loses $B_{\rm r}(M-1)$, paying the remaining $C^M$ cost and
added windows. Nonfusing rods lose no such weight and incur no such cost by
\eqref{s3n:cross-wick-intertwining}. Choose the occupied/constituent
reserve so its loss factor $\kappa_{\rm wt}<1/32$, then $L$ large enough
that
$\kappa_{\rm wt}+C_{\rm gap}L^{-3/4}<1/16$, restoring every field reserve.
Finally choose $r_*,\sigma_*$ with $C_Lr_*,C_L\sigma_*<1/16$ and
$C_{\rm Ward}(r_*+u_*)<1/8$, where $u_*\le C_{\rm prim}r_*$ bounds
the stored primitive/kernel components in the ordinary norm. Shrink $r_*$
until the single occupied reserve pays the entire contact forest.
Here $B_{\rm r}$ never weights internal Wick edges.

Interpolate actual and canonical positive shells at the same fine
spacing. A marked difference edge has both operator and fresh primitive
bounds $C_L\sigma_j^{\rm op}w$. The restored estimates give the gap and coefficient differences.
Other connected terms have at least two inputs, so the forest/contact
majorants give \eqref{s3n:ordinary-map-bounds}. Parity and the singleton
remove quadratic charged birth. Pinned bounds retain leading forcing,
one-root products and coincidence contacts.

For comparison with zero spacing, only the fresh singleton and local
current projector move. The exact identity
$e^{i\alpha q\varphi(x)}=e^{i\alpha q\varphi(a)}
 e^{i\alpha q\int_a^x\psi}$, $a\le x\le a+\delta$, has coefficient
error at most $\sum_{n\ge1}(C\rho\delta)^n/\sqrt{n!}\le C\delta$.
Also $\|\mu_{\delta,a}-\operatorname{ev}_a\|_{H^{-4}}\le C\delta$.
Inherited anchor measures and dual ports stay fixed; a pure inherited
pair has zero bulk linear output. Hence $A_{\mu,j},B_j,d_j,\Lambda_j$
have error $C_L(\sigma_j^{\rm op}+\delta)$, with $\delta\le\sigma_j^{\rm op}$.
Cauchy estimates give three current derivatives of
\eqref{s3n:microscopic-jet-distance} and six bounded derivatives of the maps.
\end{proof}

The off-graph neutral recurrence is
$R_{0,j+1}=A_{0,j}R_{0,j}+B_jy_j^2+\cdots$; its homogeneous data and
physical temperature derivatives remain part of reduction.
The complete coefficient \eqref{s3:heat-Q-value} includes this feedback,
whereas $d_j$ alone does not. On the selected weak graph parity gives $R_0=O(v^2)$ and $R_1=O(v^3)$;
physical entry and differentiated transients are proved next.

\section{Fixed physical entry and complete source reconstruction}
\label{sec:fixed-entry}
Use the temperature-independent positive cores $J_H$ and critical
records of Theorem~\ref{s3:thm:positive-core} and
Proposition~\ref{s3:prop:core-plateau}. Throughout the transfer and
slow-plane proofs, $\beta=\beta_c$, $m=m_c>0$, and $2\beta Bm^2=1$.
Write $P_H,\pi_H$ for the exact $H$-word Doob transfer and its invariant
law, $U$ for spin reversal, and $\mathcal R$ for word reversal. Constants depend
on the fixed interaction; $J_H(r)\le M/r^2$ with $M<\infty$.
Transfer norms are on $L^2(\pi_H)$; $\Pi$ denotes orthogonal projection onto the indicated span.
Put $M_{n,i}=n^{-1}\sum_{r=1}^{n}\sigma_{n(i-1)+r}$ and $M_n=M_{n,1}$;
$\operatorname{av}_i$ averages the indicated blocks. Let $\mu_J^\pm$ be the full-$J$ phases.

\subsection{The physical transfer gap}

For an end strip $C_d$ of length $d$, the conditioned stationary
two-word law has residual interaction bounded by
$D_d=\beta\sum_{r>d}(r-d)J_H(r)\le\beta M\log(H/d)$.
Its density relative to its conditional product marginals is at least
$e^{-4D_d}$. Subtracting this multiple of the product measure gives
\begin{equation}
 \E|F-G|^2\ge e^{-4D_d}\E\operatorname{Var}(F\mid C_d),
 \label{s3:direct-end-localization}
\end{equation}
for arbitrary $L^2$ functions of the respective words, and the reversed
inequality for the other word.

Any word law $\mu_H$ with
$D(\mu_H\Vert\pi_H)=O(\log H)$ satisfies, on each fixed positive-fraction
buffer and at complete $n$-block origins $i$,
\begin{equation}
 \limsup_H\operatorname{av}_i\E_{\mu_H}(|M_{n,i}|-m)_+
 \le2\E_{\mu_J^+}(M_n-m)_+\longrightarrow0\quad(n\to\infty).
 \label{s3:rare-overshoot}
\end{equation}
Indeed the word/free-word density ratio lies between $e^{-4C_H}$ and
$e^{4C_H}$, where $C_H=\beta\sum_{r\le H}rJ_H(r)=O(\log H)$.
Pasting independent $\mu_H$ words and averaging origins on the $n$-grid
costs relative entropy and seam energy per site
$O(\log H/H)$; $\|J_H-J\|_1=O(H^{-1})$. Every local limit therefore
maximizes the full-$J$ variational free energy and is Gibbs~\cite{Georgii}.
Entropy affinity gives the same conclusion for every positive-weight
buffer component. Stochastic bounds by $\mu_J^\pm$ give
\eqref{s3:rare-overshoot}; plus-state clustering makes its right side
vanish. Define $\Sigma_H=\operatorname{sgn}M_H$ with any spin-odd
$\{\pm1\}$ tie rule. Conditioning on $\Sigma_H=1$ costs $\log2$ in entropy.
The record
primitive fixes its buffer mean to $m$; monotone coupling identifies
the local limit with $\mu_J^+$.
Consequently
\begin{equation}
 \operatorname{av}_i\E_{\pi_H}|M_{n,i}-m\Sigma_H|
 \longrightarrow\E_{\mu_J^+}|M_n-m|=:e_n\longrightarrow0.
 \label{s3:ordinary-local-phase}
\end{equation}
\paragraph{The weighted cut estimate.}
For independent edges of intensities
\[
 \theta_{ij}\le[\kappa u_iu_j+\delta]/(j-i)^2+e_{j-i},\quad
 0\le u_i\le1,\quad \sum_{d\le R}de_d\le E_*,
\]
with range $R\asymp N$, suppose a dyadic interval $I$ of $N$ cuts has
mean weight at most $q_*<d<1$ and
$s=\kappa(1+\eta)d^2+\delta<1$. There is a uniformly positive
probability of a closed cut in $I$.
Let $E_k$ average on dyadic intervals of length $2^k$, and let
$A_t^-,A_t^+$ be the normalized left/right geometric averages with kernel
$(1-e^{-t})e^{-ta}$, $a\ge0$. Then
\begin{equation}
 \sum_k\int_{2^{-k-1}}^{2^{-k}}
 (\|A_t^-f-E_kf\|_2^2+\|A_t^+f-E_kf\|_2^2)\,dt/t\le C\|f\|_2^2.
 \label{s3:weighted-cut-squarefunction}
\end{equation}
Decompose $f$ into $(E_j-E_{j+1})f$. On the $k$th band,
boundary crossing for $j\ge k$ costs $C2^{(k-j)/2}$; summation by
parts against the zero-sum primitive for $j<k$ costs $C2^{j-k}$.
Young's inequality and orthogonality sum these factors. Apply this to
$f=u\ind_I$. The dyadic maximal inequality, removal of a fixed small
boundary layer, and Markov's inequality for the square function leave
$S\subset I$, $|S|\ge aN$, on which all dyadic means are at most $d$
and the square function is bounded. Inserting
$r^{-2}=\int_0^\infty te^{-tr}\,dt$ gives, simultaneously for
$1\le\ell\le N$ and $z\in S$,
\[
 \sum_{r=\ell}^Nr^{-2}\sum_{a=0}^{r-1}f(z-a)f(z+r-a)
 \le(1+\eta)d^2\log(N/\ell)+C.
\]
The ranges $t<1/N$, $t>1/\ell$ cost $O(1)$; in between use
$(A_t^-f)(A_t^+f)\le(|A_t^-f|^2+|A_t^+f|^2)/2$ and
$|A_t^-f|^2\le(1+\eta)|E_kf|^2+(1+\eta^{-1})|A_t^-f-E_kf|^2$.
For closed-cut events $A_z$, independence therefore gives
\[
 \frac{\Pr(A_z\cap A_w)}{\Pr(A_z)\Pr(A_w)}
 \le C[N/(1+|z-w|)]^s.
\]
Edges outside $I$ cost $O(1)$ on $S$; the $e_d$ cost $E_*$.
The normalized inverse-probability count
$|S|^{-1}\sum_{z\in S}\ind_{A_z}/\Pr(A_z)$ has mean one and bounded
second moment since $s<1$. The second-moment inequality proves the claim.

\begin{theorem}[Two-step even gap]
\label{s3:thm:two-step-gap}
At all sufficiently late records, for some $\rho<1$,
\begin{equation}
 \|P_H^2\|_{\{Uf=f,\ \pi_Hf=0\}}\le\rho,
 \qquad \|P_H^k-\Pi_1\|_{U=1}\le\rho^{\lfloor k/2\rfloor}.
 \label{s3:two-step-gap}
\end{equation}
\end{theorem}
\begin{proof}
Otherwise choose real centered even unit $f_H$ with
$\|P_H^2f_H\|_2\to1$, and put $g=P_Hf_H$, $h=P_H^2f_H$.
The pair errors $\E|g(X_0)-f_H(X_1)|^2=1-\|g\|_2^2$ and
$\E|h(X_0)-g(X_1)|^2=\|g\|_2^2-\|h\|_2^2$ vanish.
Equation~\eqref{s3:direct-end-localization} approximates $g$ by centered
even $F,G$ on the first and last $H/8$ strips, with variances tending
to one and $\|F-G\|_2\to0$. The seam bound
$P_H(x,y)/\pi_H(y)\le e^{4C_H}$ gives $\|g\|_\infty\le e^{2C_H}$.
Thus $g^2\pi_H/\|g\|_2^2$ has entropy $O(\log H)$; the normalized
squared laws of $F,G$ approach it in total variation.

For two replicas, condition on unordered pairs of $n$-blocks
$\{a_i,b_i\}$, and write their orientations as
$(a_i+b_i)/2\pm\eta_i(a_i-b_i)/2$. The exact orientation coupling is
$K_{ij}=2\beta d_i^{\mathsf T}J_{ij}d_j$, where $d_i=(a_i-b_i)/2$.
For either global alignment $\varepsilon=\pm1$, set
$u_i^\varepsilon=|M_{n,i}(\sigma)-\varepsilon M_{n,i}(\tau)|/2$.
For $D=j-i\ge2$, $R=\lceil H/n\rceil+1$, the actual core estimate gives
\[
 2|K_{ij}|\le[4\beta Bu_i^\varepsilon u_j^\varepsilon+\delta_n]/D^2
       +C\beta B(D^{-3}+R^{-2}),\quad
 \delta_n=C\beta B\sup_{r\ge n}|r^2J_r/B-1|\to0.
\]
Adjacent blocks have a fixed finite cost at fixed $n$; the remaining
error has bounded first moment. Choose
$m/2<q_*<d<m/\sqrt2=(4\beta B)^{-1/2}$ and then $n$ so large that
the cut exponent is below one. The inequality
\[
 \min_{s=\pm1}\operatorname{av}_i|x_i-sm|/2
 \le m/2+\operatorname{av}_i(|x_i|-m)_+
\]
and \eqref{s3:rare-overshoot}--\eqref{s3:ordinary-local-phase} show that
at least one alignment has mean $u_i^\varepsilon\le q_*$ on a
fixed middle interval between the strips, except on a set of
arbitrarily small squared-law mass.

The signed bond identity
$e^{K\eta_i\eta_j}=e^{|K|}[(1-p)+p\ind_{\eta_i\eta_j=\mathrm{sgn}K}]$,
$p=1-e^{-2|K|}$, dominates open bonds by the independent intensities
$2|K|$. Given compatible bonds each component has two fair orientations.
The cut estimate supplies a closed cut with probability at least $c_n$,
uniformly in the conditioned orientations. Flip all components to its
left. For $\mathsf F=F(\sigma)-F(\tau)$ and
$\mathsf G=G(\sigma)-G(\tau)$, spin-evenness handles either alignment;
the mixed term cancels and yields
\[
 2\E(\mathsf F-\mathsf G)^2\ge
 c_n\E[\ind_{\mathcal G_+\cup\mathcal G_-}(\mathsf F^2+\mathsf G^2)].
\]
The left side tends to zero; the squared-law bounds keep the right
side positive. Iterating the contraction gives the second assertion.
\end{proof}

\paragraph{From phase mass to operator norm.}
Let $\epsilon_H$ be the probability of a phase change across one word;
primitive convergence gives $\epsilon_H\to0$. In spin-orbit coordinates
$P((x,s),(y,t))=\bar P(x,y)(1+st a_{xy})/2$, $|a_{xy}|\le1$.
Its odd kernel $A$ satisfies
$0\le\bar P^k-A^k\le2\bar P^k$ and total mass at most $2k\epsilon_H$.
If $0\le D\le MK$, $\|K-\Pi_1\|\le e$, and $\langle1,D1\rangle=\delta$,
truncate unit test vectors at height $\Lambda$. Bounded parts cost
$\Lambda^2\delta$; tails cost $2M/\Lambda+2Me$. Optimize $\Lambda$ to obtain
\begin{equation}
 \|P_H^k-\Pi_{1,\Sigma_H}\|
 \le C\rho^{\lfloor k/2\rfloor}+C(k\epsilon_H)^{1/3}.
 \label{s3:full-phase-compression}
\end{equation}

\subsection{Frozen slow variables and physical preparation}

\begin{proposition}[Microscopic weights and the slow plane]
\label{s3:prop:microscopic-weight}
\label{s3:thm:microscopic-slow-plane}
For $|w_i|\le H^{-1}$,
\begin{equation}
 \sup_w\left\|\sum_iw_i\sigma_i-m(\sum_iw_i)\Sigma_H\right\|_2\to0.
 \label{s3:arbitrary-weight-phase}
\end{equation}
The one-site Doob transfer $\mathsf T_H$, with $\mathsf T_H^H=P_H$,
has a commuting rank-two spectral projection $E_H\to\Pi_{1,\Sigma_H}$,
even part $\Pi_1$, and
\begin{equation}
 \|\mathsf T_H^n(1-E_H)\|\le C\rho_{\rm f}^{\lfloor n/(k_*H)\rfloor},\qquad 0<\rho_{\rm f}<1.
 \label{s3:microscopic-fast-gap}
\end{equation}
Its odd eigenvalue satisfies $0<\nu_H<1$, $\nu_H^H\to1$. For multiplication
$\mathsf S$ by the last spin, bounded dual
slow maps $V,W$, with $WV=I$ and $VW=E_H$, give
\begin{equation}
 W\mathsf T_HV=e^{-y_R/R}e^{(y_R/R)\sigma_x},\quad
 W\mathsf SV=Z_H\sigma_z,\quad y_R=-R\log\nu_H/2,
 \quad Z_H\to m.
 \label{s3:microscopic-current}
\end{equation}
\end{proposition}
\begin{proof}
Primitive convergence on $3H$ aligns all sliding $H$-windows, so
$Z_{H,i}=\E\sigma_i\Sigma_H\to m$ uniformly in $i$.
Write $c_H(d)$ for the stationary core two-point function. Condition
outside two radius-$r$ windows at distant sites. Their
mutual interaction is $O(r^2/d^2)$; the core's excess over $J$ is
absorbed by a uniform field $O(H^{-1})$. Plus-boundary comparison,
then $H,d,r\to\infty$ in that order, gives
$\lim_{d_0\to\infty}\limsup_H\sup_{d\ge d_0}c_H(d)\le m^2$.
For $K_{ij}=c_H(j-i)-Z_{H,i}Z_{H,j}$, positive row parts thus vanish
after division by $H$. Signed row sums vanish because
$M_H-(\E|M_H|)\Sigma_H\to0$ in $L^1$. Since $|x|=2x_+-x$,
$\sup_iH^{-1}\sum_j|K_{ij}|\to0$, proving the weight bound.

Choose a fixed $k_*$ and then sufficiently late $H$ so that
\eqref{s3:full-phase-compression} makes
$\|P_H^{k_*}-\Pi_{1,\Sigma_H}\|<1/16$.
The Neumann resolvent series about this orthogonal projection on
$|z-1|=1/4$ and $|z|=1/2$ gives a rank-two Riesz projection $E_H$
and uniform complementary-power bounds for $P_H^{k_*}$.
Continuous perturbation along the same contours preserves one
dimension of each spin parity.
Since $P_H^{k_*}=\mathsf T_H^{k_*H}$, $E_H$ commutes with
$\mathsf T_H$ and spin reversal. Its even line is constant, and its
odd line has a real $\mathsf T_H$-eigenvalue $\nu_H$.
The fixed small projection error gives $\|E_H\Sigma_H\|_2\ge c>0$.
Phase persistence and commutation imply
\[
 (\nu_H^H-1)E_H\Sigma_H=E_H(P_H-I)\Sigma_H\longrightarrow0,
 \qquad
 \|P_H^k-E_H\|\le C\rho_{\rm f}^{\lfloor k/k_*\rfloor}
                         +C|\nu_H^{Hk}-1|.
\]
Thus $\nu_H^H\to1$. Combining the second bound with
\eqref{s3:full-phase-compression}, first let $H\to\infty$ at fixed
$k$ and then $k\to\infty$. This proves
$\|E_H-\Pi_{1,\Sigma_H}\|\to0$, controlling both dual phase lines.
The complementary-power bound and contraction of $\mathsf T_H$
give \eqref{s3:microscopic-fast-gap}.
A one-site shift changes $M_H$ by at most $2/H$, so primitive
convergence gives $(\mathsf T_H-I)\Sigma_H\to0$ in $L^2$.
Applying $E_H$ shows $\nu_H\to1$ and hence $\nu_H>0$ at late records;
primitivity gives $\nu_H<1$. In parity coordinates $W\mathsf SV$ has positive
off-diagonal entries $a_H,b_H\to m$. A reciprocal rescaling of dual
odd vectors makes both $\sqrt{a_Hb_H}=Z_H$; the Hadamard transform gives
the phase form. If $A_w$ denotes multiplication by $\sum_iw_i\sigma_i$,
the weight estimate bounds $(1-E_H)A_wE_H$ and its adjoint;
multiply by the original $\Sigma_H$, with $\Sigma_H^2=1$,
to obtain the second column.
\end{proof}

\begin{lemma}[Fixed reflection and centered sources]
\label{s3:lem:fixed-reflected-transfer}
The preparation admits fixed real dual slow bases with
$W_B=V_B^*\mathcal R$, $W_BV_B=I$. Its physical source current remains $Z_H$.
\end{lemma}
\begin{proof}
For an allowed one-site shift $x\to y$, use
$T_\beta(x,y)=\exp[\beta(A_{\rm new}+A_{\rm old})/2]$.
Since $A_{\rm new}-A_{\rm old}=\mathcal E(y)-\mathcal E(x)$ for the
internal word energy $\mathcal E$, half-word endpoint energies
recover the physical transfer, and $T_\beta^{\mathsf t}=\mathcal R T_\beta\mathcal R$.
Fix Perron data $r_c,\lambda_c$ at $\beta_c$ and
$\pi_c(x)\propto r_c(x)r_c(\mathcal R x)$; then
$\mathsf P_\beta=\lambda_c^{-1}D_{r_c}^{-1}T_\beta D_{r_c}$ obeys
$\mathsf P_\beta^*=\mathcal R\mathsf P_\beta\mathcal R$ on this fixed space.

On anti-palindromic ties, an odd $\{\pm1\}$ proxy cannot be reversal even.
Nevertheless
$\|(\mathcal R-I)\Sigma_H\|_2\le2\sqrt{\pi_H(M_H=0)}=o(1)$.
The bounded, reversal-even, spin-odd function $(\Sigma_H+\mathcal R\Sigma_H)/2$
has norm $1+o(1)$. Thus
$\langle\mathcal R f,f\rangle\ge(1-o(1))\|f\|^2$ on $\operatorname{ran}E_H$;
orthonormalization gives $W_B=V_B^*\mathcal R$. This symmetrization is used
only here; the preceding square-one identities use $\Sigma_H$.

Split each microscopic source equally between entrance and departure
of its spin; the word gauge telescopes, retaining missing endpoint
halves. Put $D=W_B\mathsf P_cV_B=\operatorname{diag}(1,\nu_H)$ and $m_0=H-1$.
The shift identity $\mathsf P_c^{m_0}e^{t\mathsf S_{\rm first}}
=e^{t\mathsf S_{\rm last}}\mathsf P_c^{m_0}$ implies, for
$A=W_B\mathsf S_{\rm last}V_B=\left(\begin{smallmatrix}0&a_+\\a_-&0\end{smallmatrix}\right)$,
$a_-=a_+\nu_H^{m_0}$ and
\begin{equation}
 W_Be^{t\mathsf S_{\rm last}}\mathsf P_c^{m_0}V_B
 =D^{m_0/2}(\cosh t\,I+Z_H\sinh t\,\sigma_x)D^{m_0/2},
 \quad Z_H=\sqrt{a_+a_-}.
 \label{s3:centered-source-guard}
\end{equation}
The fine-cell reference is
$D_{\rm ph}^{1/2}(\cosh\xi_i I+Z_H\sinh\xi_i\sigma_z)D_{\rm ph}^{1/2}$.
Its difference from
$e^{Z_H\xi_i\sigma_z}$, including contact curvature $(1-Z_H^2)I$,
stays in the ordinary coefficient. Reflection transposes the source
word and exchanges endpoints; its fast form need not be positive.
\end{proof}

\paragraph{Entry in the common norm.}

\begin{proposition}[The rough short-covariance network]
\label{s3:prop:initial-network}
At $R=kH$, the actual short covariance differs from its canonical
counterpart by an exact network of incidence norm $O_k(E(H/16))$.
Its charged potential obeys a unit-window $CkE(H/16)(1+r^2)^{-1}$ bound.
\end{proposition}
\begin{proof}
For gradient covariances $\Gamma_H,\Gamma_H^0$,
$\|\Gamma_H-\Gamma_H^0\|_1\le CE(H/16)/H$.
The short heat filter has bounded $\ell^1$ norm, so its difference
$\Delta\Gamma$ has the same bound. For every finite spin path the
Gaussian-factor ratio is exactly
\[
 \exp[-\beta BN\Delta\Gamma(0)/2]
 \prod_{i<j}e^{-\beta B\Delta\Gamma(j-i)\sigma_i\sigma_j}.
\]
The absolute pair strength incident on an $R$-block is
$\epsilon=O(kE(H/16))$.
Expanding pair factors around one and summing spanning trees gives, for $s$ blocks,
$e^{C\epsilon s}s^{s-2}(C\epsilon)^{s-1}/(s-1)!$.
Thus the occupied-support sum is $O(\epsilon)$, with the diagonal
scalar retained and no interval hull assigned to long edges. For the primitive, the identity
$(1-e^{-R^2\lambda})/\lambda=\int_0^{R^2}e^{-t\lambda}dt$
gives a Green kernel bounded by $CR^2w_R$, where
$w_R(n)=R^{-1}(1+|n|/R)^{-2}$; the cutoff correction is Schwartz.
Since the original defect is bounded by
$CE(H/16)H^{-1}(\ind_{\{0\}}+w_H)$ and $w_R*w_H\le Cw_R$, its potential is
bounded by $CkE(H/16)(1+|n|/R)^{-2}$.
Ward normalization (Theorem~\ref{s3n:thm:common-map}) converts the
derivative pair, retaining contacts and endpoint halves. Source
derivatives insert norm-one spins; temperature derivatives use a fixed disk.
\end{proof}

\begin{proposition}[Finite physical preparation]
\label{s3:prop:finite-physical-preparation}
\label{s3:lem:entry-fock-tree}
Every fixed $J\in\mathcal J_B$ enters the common norm at finite $H_0$,
on an open temperature interval containing $\beta_c$.
Core, Hilbert space, slow bases, source convention, covariance split
and extraction chart are fixed on that interval. Bulk activity,
charge deviation and coefficients are arbitrarily small; physical
endpoints are finite analytic external modules. All required source
coefficients have two temperature derivatives; the full head and tail
base has zero temperature derivative.
\end{proposition}
\begin{proof}
For $R=kH$ the exact sourced block kernel is
\begin{equation}
 K(u)f(x)=\E_x\!\left[e^{\sum_{i=1}^{R}\sigma_i
                  \int_{(i-1)/R}^{i/R}u(t)dt}f(X_k)\right].
 \label{s3:entry-exact-sourced-kernel}
\end{equation}
With $\mathfrak M(u)=e^{\|\Re u\|_{L^1}}$, primitive integration by
parts and $H^2\hookrightarrow C^1$ bound its stationary mean error
from the constant-phase kernel by
$C\mathfrak M(u)\eta_H(k)(1+\|u\|_{H^2})$, where
$\eta_H(k)\to0$ at fixed $k$.
The absolute error is bounded by $2\mathfrak M(u)P_H^k$.
The truncation proof of \eqref{s3:full-phase-compression}, with its
rank-two phase projection, converts this to operator error of exponent
$1/3$. For $u=z+i\psi^{\rm sh}$, with $\psi^{\rm sh}$ a real short
Gaussian current, the $\psi^{\rm sh}$ factor has modulus one and the envelope is
\[
 Ce^{C\|z\|_{H^2}}\{\rho_{\rm f}^{\lfloor k/k_*\rfloor}
       +t_H(k)(1+\|z+i\psi^{\rm sh}\|_{H^2})^{1/3}\},\qquad t_H(k)\to0.
\]
Cauchy estimates give the current and source jets.

Expand the two terms separately and merge adjacent occupied blocks.
For separated rods, the logarithmic variogram minus its heat average
gives nuclear covariance rows $C(1+r^2)^{-1}$. A positive spectral
extension by a smooth partition over $2\pi\mathbb Z$ agrees at lattice
sites; its Schwartz interpolation kernel reproduces polynomials,
preserving these bounds off the compact core. Coincident bounds use
$H^2$ moments. Only primitive-error blocks require polynomial Gaussian
moments: on their set $S$, the covariance has operator norm $M_k$ and
trace at most $M_k|S|$. The bound
$(1+s)^{1/3}\le C\epsilon^{-1/6}e^{\epsilon s^2}$,
$\epsilon<(8M_k+8)^{-1}$, costs $C_k^{|S|}$.
The Fock embedding and rooted sum of Section~\ref{s3n:extraction} then
cost $C$ per transfer-error block, independently of $k$, and no
inactive-block determinant. Full and half charge normalization cost
$k^a$, since short variance is at most $\log k+C$.

Average seam insertions $E_c+(1-E_c)$, $E_c=E_H$, over the stationary
renewal with length law $a_R$ uniform on $\{R,\ldots,2R\}$ and mean
$\mu_R=3R/2$. Its renewal mass satisfies
\[
 u_R(n)\le2/R\ (n\ge R),\qquad
 |u_R(n)-\mu_R^{-1}|\le(C/R)e^{-cn/R}\ (n\ge3R).
\]
The renewal equation gives the first bound. Seven convolutions have
mass $\ge c/R$ on $[8R,12R]$; thus the countdown chain minorizes its
stationary law after $12R$ steps. Contraction and $3R$ smoothing give
the second.
For ordered selected cells $[a_i,b_i)$ the exact intensity is
$\mu_R^{-1}\prod_i a_R(b_i-a_i)\prod_i u_R(a_{i+1}-b_i)$.
Sum consecutive occupied strings first, and split each remaining gap
as $u_R(g)=\mu_R^{-1}[1+h_R(g)]$. The linked part has
$|h_R(g)|\le Ce^{-cg/R}$; its zero contact is already in the string sum.
Choose fixed $Q$ so the exponential pays hull and holonomy costs in
units $H_0=QR$. Rooted chains sum geometrically, with one small rod
factor per attachment. Symmetric padding and local closure merge short
gaps with their exact propagators. Stationarity preserves translation,
reflection and, since $\mu_R^{-1}\sum\ell a_R(\ell)=1$, the physical
current. Partial endpoint cells remain external. The renewal average is exact.

Combining the two-type forest bound with the rough network gives
\begin{equation}
\begin{gathered}
\epsilon_{\rm bulk}\le C_Qk^a\rho_{\rm f}^{\lfloor k/k_*\rfloor}
       +C_{Q,k}\omega_H(Q,k),\qquad \omega_H(Q,k)\to0,\\
 \sigma_{H_0L^j}^D\le\sigma_{H_0L^j}^{\rm G}
                  +(C+o(1))(Qk)^{-1}L^{-j}.
\end{gathered}
 \label{s3:separated-entry-bound}
\end{equation}
The second bound follows from Section~\ref{s3:heat-tail-base},
using $B^{-1}\sum_{n\le H}n^2J_H(n)=O(H)$.
Choose $Q$, then $k$ for the mixing and independent $1/(Qk)$ errors,
then a record $H$ for $\omega_H$ and the tail distance. The physical
charge $4\beta_cBZ_H^2\to2$ and activity
$-(Qk/2)\log\nu_H^H\to0$ meet their reserves simultaneously.

Freeze all preparation data. Nearby transfers have
$O(|\beta-\beta_c|)$ excursions from $E_c$; these remain ordinary
coefficients. Finite-matrix analyticity gives a complex temperature
disk whose gap moments fit the reserve and hence two Cauchy derivatives.
Extraction defines $Z_0(\beta)$ with charge $4\beta BZ_0(\beta)^2$.
The fixed base retains its head memory (Sec.~\ref{s3:heat-tail-base});
any preliminary prefix is fixed before entry.
\end{proof}

\begin{proposition}[Finite physical endpoints]
\label{s3:prop:endpoint-module}
Physical free endpoints enter as finite analytic modules; their norms
need not be small and occur at most once per endpoint.
\end{proposition}
\begin{proof}
Use word-transfer Perron data $r_c$, eigenvalue $\lambda$, and left
vector $\ell=c_H\mathcal R r_c$ normalized by $\ell^{\mathsf T}r_c=1$.
For the positive half-word vector $v^f$, put $e=v^f/r_c$,
$a=\langle1,e\rangle_{\pi_c}>0$, $\widehat e=e/a$. Since $\widehat e$ is even of mean one,
$\ell_e(f):=\langle\mathcal R\widehat e,E_cf\rangle_{\pi_c}=\langle1,f\rangle_{\pi_c}$.
Retain the missing first transition, endpoint source halves and
Perron scalar in the transfer word between $\mathcal R\widehat e$ and $\widehat e$.
For complete blocks,
\begin{equation}
 c_H\mathcal Z^f/(a^2\lambda^{kT-1})
 =(\ell_eK_1+\Gamma_1)K_2\cdots K_{T-1}(K_T1+\Delta_T),
 \label{s3:entry-exact-boundary}
\end{equation}
where $\Gamma_1=\langle \mathcal R\widehat e,G_1\cdot\rangle-\ell_eK_1$,
$\Delta_T=K_T(\widehat e-1)$, and $G_1$ omits that first transition.
Insert $E_c+(1-E_c)$. With bulk excursion bound $d$, anchored strings have
norm $C(1+\|\widehat e\|_2)d^{\ell-1}$; a wholly fast span has
$C(1+\|\widehat e\|_2)^2d^{T-2}$. Their anchored exponential sums
converge. At zero source the endpoint is even of mean one, so its
slow component is exactly the free vector and mixed fast zero jets
vanish. Forests are linear in each external insertion; row sums multiply
its finite root constant without weighting the distance between roots.
The fixed change $(1,1)^{\mathsf T}\mapsto(1,-1)^{\mathsf T}$ preserves
the current and scalar. The same temperature disk applies.
The free/periodic energy difference is
$O(\sum_n\min(n,N)J_n)=O(1+\log N)$; their thermodynamic pressures
agree at every fixed depth with identical source weights.
\end{proof}

\subsection{Reduction and the critical separator}
\label{sec:reduction-sources}
\label{s3:complete-reduction-coefficients}
Apply the common heat map (Sec.~\ref{s3n:extraction}) to the fixed preparation.

Write the common map as $(f_j(x,R),g_j(x,R))$, with finite-dimensional
center $x$ and ordinary coordinate $R$. After the limiting shear,
$D_xf_j=I+O(\epsilon)$, $D_Rf_j,D_xg_j=O(\epsilon)$ and
$\|D_Rg_j\|\le\kappa<1$. All six current derivatives are bounded on
fixed domains. Extend the center variables by a smooth cutoff outside
a smaller chart to give global near-identity inverses, leaving physical
prefixes unchanged. Starting with $\Phi_0=0$, define
$f_j^\Phi(x)=f_j(x,\Phi_j(x))$ and
$\Phi_{j+1}\circ f_j^\Phi=g_j(x,\Phi_j(x))$.
Differentiation gives
$D\Phi_{j+1}\circ f_j^\Phi=(g_x+g_RD\Phi_j)(f_x+f_RD\Phi_j)^{-1}$.
At order $r\le6$ the highest graph jet has coefficient
$(\kappa+C\epsilon)(1-C\epsilon)^{-r}<1$; induction bounds the jets.
Transverse errors contract by $\kappa_0<1$, forcing center errors
$O(\epsilon\kappa_0^j)$. Pulling the $N$th center coordinate back by
$(f_0^\Phi)^{-1}\cdots(f_{N-1}^\Phi)^{-1}$ gives consecutive
phase differences $O((\kappa_0K)^N)$, where $K$ bounds center derivatives and
inverses. Shrink the chart so $\bar\kappa K^9<1$. Chain-rule estimates
through order three are bounded by $CN^C\kappa_1^NK_0^{7N}$, for
$\kappa_0<\kappa_1<\bar\kappa$, $K_0<K$; hence the phase converges in $C^3$.
The reduced shadow and its first two temperature derivatives track
the full state geometrically. Only center maps are inverted.

For the neutral quadratic graph, let
$\lambda_j(q)=\exp[-q-(1+q/b)a_j]$ be the singleton multiplier and
$A_j,B_j$ be the neutral linear operator and quadratic birth. Then
\begin{equation}
 T_{j+1}=\lambda_j^{-2}(A_jT_j+B_j),\qquad
 R_j=\mathcal U(j,0)R_0+\sum_{i<j}\mathcal U(j,i+1)f_i.
 \label{s2:memory}
\end{equation}
Here $f_i$ is the nonlinear forcing and $\mathcal U(j,i)=A_{j-1}\cdots A_i$.
The first identity follows from graph invariance; its shifted charge
argument changes only quartic terms. On a reserved
disk $\|\lambda_j^{-2}A_j\|<\widehat\kappa<1$. Subtracting the
canonical resolvent $T_*=(\lambda_*^2I-A_*)^{-1}B_*$ and using Cauchy
estimates bounds its first three derivatives by
\begin{equation}
 \overline\sigma_j=C\rho^{j+j_b}+C\sigma_j^{\rm op}
       +C\sum_{-j_b\le i<j}\rho^{j-1-i}\sigma_i^{\rm op}\longrightarrow0,
 \quad \widehat\kappa<\rho<1.
 \label{s3:actual-coefficient-distance}
\end{equation}
Here $j_b$ is the fixed preliminary length. If
$R_{j+1}=A_jR_j+B_jv_j^2+o(v_j^2)$, $A_j\to A_*$,
$B_j\to B_*$ and $v_{j+1}/v_j\to1$, division by $v_j^2$ gives a
contracting affine recurrence, so $R_j/v_j^2\to(I-A_*)^{-1}B_*$,
without an asserted rate.

\begin{theorem}[The actual reduced map]
\label{s3:thm:actual-reduced-map}
After a fixed finite preliminary prefix, absorbed into preparation,
\begin{align}
 u_{j+1}&=u_j-k_j(u_j,v_j)v_j^2,\notag\\
 v_{j+1}&=v_j\exp[-u_j-(1+u_j/b)a_j+v_j^2h_j(u_j,v_j)],
 \label{s3:proved-reduced-map}\\
 |k_j-1|&\le C(|u|+|v|+\overline\sigma_j),\quad
 0<k_-\le k_j\le k_+,\quad |Dk_j|+|h_j|+|Dh_j|\le C.
 \label{s3:proved-coefficient-bounds}
\end{align}
The physical charge satisfies
\begin{equation}
 q_j=u_j+\zeta_j(u_j,v_j)+t_j,\quad
 |\zeta_j|+|\partial_u\zeta_j|\le Cv_j^2,\quad
 |\partial_v\zeta_j|\le C|v_j|,\quad
 |\partial_\beta^rt_j|\le C_J\theta^j\ (r\le2).
 \label{s3:proved-physical-chart}
\end{equation}
There is no profile-only charge shift. Neutral source families are
$O(v_j^2)+O_J(\theta^j)$ and charged families
$O(v_j)+O_J(\theta^j)$, including every source-dependent scalar.
\end{theorem}
\begin{proof}
Write $f_j,g_j$ for the reduced components. Parity makes the neutral
graph quadratic; singleton closure removes the proper charged linear
term. Taylor division extends
$(u-f_j)/v^2$ and
$[\log(g_j/v)+u+(1+u/b)a_j]/v^2$ with bounded first derivatives.
The limiting complete quadratic coefficient is positive. For the heat
map Proposition~\ref{s3:thm:heat-coefficient} gives
$\gamma_*=4b^2e^{-\gamma_E}$, so $v=\sqrt{\gamma_*}y$ at first order.
Current extraction and the ordinary resolvent both enter this
coefficient. A bounded convergent quadratic change $c_j(q)y^2$
alters the critical coefficient by
$c_{j+1}(0)\lambda_j(0)^2-c_j(0)\to0$; it cannot alter its limit.
Equation~\eqref{s3:actual-coefficient-distance} controls the other jets.
Choose $j_b$ to put homogeneous memory below the drift reserve and
prepare for this fixed prefix. Taylor bounds and stable tracking finish the proof.
\end{proof}

\begin{proposition}[Positive separator and its derivative]
\label{s2:prop:critical-existence}
\label{s3:thm:critical-separator-c1}
For sufficiently small positive activity and a sufficiently late fixed
starting index, there is a unique positive weak survivor with
$u_j\downarrow0$. Its initial charge is a continuous graph
$u=\Gamma_j(v)$, $C^1$ at every $v>0$. Above it the limiting charge is
positive; below it the charge becomes negative in finite time.
\end{proposition}
\begin{proof}
On a small box the derivative signs are $f_u,g_v>0$, $f_v,g_u<0$,
so the map is competitive. At $u=U$, $v<c_0U$, and
$\sup|a_j|<U/8$, the bootstrap $u\ge U/2$ gives an activity ratio
$\le e^{-U/4}$ and total charge loss $\le4k_+c_0^2U<U/4$.
This gives an open ordered trap. Nonnegative-charge prefixes obey
$\sum v_j^2\le U/k_-$ and $v_j\le2\sqrt{U/k_-}$, and hence remain confined
before passage to an infinite horizon. Finite-horizon admissible charges
are closed upper intervals; their lower endpoints have final charge
zero and increase to a survivor. A positive limiting charge enters
an open ordered trap and would allow a smaller initial charge; hence
the threshold limit is zero. Two survivors satisfy
$d_{j+1}\ge(1-Cv_j^2)d_j$, whose product is positive. They cannot both
have zero limit. Openness of the two exit sets gives continuity.

Put $w=\log v$, $\gamma_j(w)=\Gamma_j(e^w)$, and use $f,g$ for the
map in $(u,w)$. In the weak box,
$f_u,g_w=1+O(v^2)>0$, $-f_w\asymp v^2$, $-g_u\asymp1$ and
$\det DF_j>0$. The finite-horizon threshold $u_N=0$, with a
nonnegative-charge prefix, defines a $C^1$ graph $\gamma_{j,N}$ by
the implicit-function theorem. Its slopes $\ell_{i,N}$ obey backward
Riccati iteration with terminal value $\ell_{N,N}=0$. With coefficients
held on a fixed critical orbit, zero-terminal iterations increase with $N$ to
\begin{equation}
 \ell_j=\frac{\ell_{j+1}g_w-f_w}{f_u-\ell_{j+1}g_u},\quad
 0\le\ell_j\le C\sum_{i\ge j}v_i^2\to0,
 \qquad n_j(T)=T_u-\ell_jT_w.
 \label{s3:stable-riccati}
\end{equation}
Indeed the fractional map is increasing, and its derivative at
terminal slope $s\ge0$ satisfies
\begin{equation}
 0<\frac{\det DF_i}{(f_u-sg_u)^2}\le1+Cv_i^2.
 \label{s3:riccati-slope-lipschitz}
\end{equation}
Since its denominator is at least $f_u\ge1-Cv_i^2$, any such
zero-terminal iteration $s_{i,N}$ satisfies
\[
 s_{i,N}\le(1+Cv_i^2)s_{i+1,N}+Cv_i^2,
 \qquad 0\le s_{i,N}\le C\sum_{r=i}^{N-1}v_r^2.
\]
The products of coefficients are bounded by $\exp(C\sum v_i^2)$.
This proves the asserted bound for the fixed-orbit Riccati limit.
On a threshold orbit, charge telescoping gives at each intermediate
time $M$ the stronger bound
$\ell_{M,N}\le Cu_{M,N}\le Cu_M^c$; the final inequality follows
from competitive comparison with the critical orbit at the same
initial $w$. No ordering of their derivatives is used.

On a compact initial $w$-interval, Dini's theorem gives uniform
convergence of $\gamma_{j,N}$ to $\gamma_j$ and of $u_M^c\downarrow0$.
At fixed $M$, the threshold orbits and the coefficients of their
Riccati maps through time $M$ converge uniformly to the critical ones.
Equation~\eqref{s3:riccati-slope-lipschitz} transports the remaining
$Cu_M^c$ terminal-slope error to time $j$ with uniformly bounded
amplification. First taking $N\to\infty$ at fixed $M$, then
$M\to\infty$, proves uniform convergence of the threshold derivatives
to the fixed-orbit Riccati limit. Hence $\gamma_j$ is $C^1$ and
$\Gamma_j'(v_j)v_j=\ell_j$. Positive Riccati denominators also give
the $C^1$ pullback through any admitted fixed finite prefix.
\end{proof}

\begin{lemma}[Critical signal]
\label{s2:lem:signal}
On the positive zero-limit branch,
\begin{equation}
 u_j=\sum_{i\ge j}k_iv_i^2,\quad v_j^2=o(u_j),\quad
 q_j\sim u_j,\quad q_{j+1}/q_j\to1,\quad \log q_j=o(j).
 \label{s2:critical-signal}
\end{equation}
\end{lemma}
\begin{proof}
Since $v_{j+1}/v_j\to1$, comparison with arbitrarily long fixed
windows in the backward sum gives $u_j/v_j^2\to\infty$. The remaining
claims follow from the map and the geometric reconstruction error.
\end{proof}

The exact normal multiplier is
$M_j=f_u-\ell_{j+1}g_u>0$, so
$n_{j+1}(DF_jT)=M_jn_j(T)$ and $\Pi_N=\prod_{j<N}M_j$.
A tangent with $n_0(T)=0$ has bounded $T_{w,j}$ because its multiplier
$g_w+g_u\ell_j=\det DF_j/M_j$ is positive and at most $1+Cv_j^2$; hence
$T_{u,j},T_{v,j}\to0$.

\subsection{Complete source response}
\label{s3:complete-source-section}

\begin{proposition}[Microscopic source modules]
\label{s3:prop:microscopic-source-norm}
For finite microscopic profiles $f,g$, the one- and two-mark entry
norms are at most $C_J\|f\|_1$ and $C_J\|f\|_1\|g\|_1$, including
two temperature derivatives. Contacts and source scalars are retained.
\end{proposition}
\begin{proof}
Each spin multiplier has norm one. The multinomial identity bounds mixed
coefficients by $\|f\|_1^m\|g\|_1^n e^{|t|\|f\|_1+|s|\|g\|_1}$.
Differentiate before the renewal and Gaussian expansions, retain the
original roots, and apply the common rooted bound on the temperature
disk. For coincident marks the centered
factor of \eqref{s3:centered-source-guard} contributes the additional
contact $(1-Z_H^2)f_ig_iI$. Distinct atoms remain distinct. Rooted maps
are linear in one-mark modules and affine in the two-mark module,
with its bilinear one-mark term.
\end{proof}

The heat derivative covariance is positive with zero zero-mode,
bounded total variation and localized nuclear rows
$\|Q_{IJ}\|_1\le C(1+|I-J|)^{-2}$ on $E_I=H^4(I)$.
Section~\ref{s3n:extraction} supplies its uniform operator/trace bounds
and bounded occupation-to-field translations. Factor the exact charge
phase from each rod entry. For the remaining analytic tensor $F_X$ on
occupied blocks $X$, use
\[
 r_A=\sup_I\sum_{X\ni I}e^{A|X|}\|F_X\|_{a;X},\qquad
 \|F\|_{a;X}=\sup_\psi|F(\psi)|e^{-a\sum_{I\in X}\|\psi_I\|_{E_I}^2}.
\]
Choose fixed field radii and $a$ with
$1/(2\rho^2)\le a<(32C_Q)^{-1}$, including cover multiplicity.
Then the Fock norm dominates this entire norm by
$x^n/\sqrt{n!}\le e^{x^2/2}$. Supports are disjoint and holes carry
exact propagators. Only the source-independent scalar $c^0$ is external;
source scalars remain even coordinates in $r_A$.

\begin{theorem}[Complete signed positive-source response]
\label{s3:thm:network-positive-response}
On the reserved current disk of Theorem~\ref{s3n:thm:common-map} and a
fixed complex source domain $\Re\eta\ge\eta_0>0$, small activity and
network norm give holomorphic complete pressure.
Writing $r_0,r_1$ for even/odd network norms,
\begin{equation}
 \widehat P=c^0+\eta+F,\qquad
 |\partial_\eta^rF|\le C_r[r_0+(|y|+r_1)^2]\quad(r=0,1,2).
 \label{s3:network-positive-pressure}
\end{equation}
The complete response and curvature have bounded derivatives in
absolute current coordinates $(p,y,R)$; their charged derivatives are
$O(|y|+r_1)$. The constants are uniform over the admitted profiles.
\end{theorem}
\begin{proof}
Divide by $e^{\eta T}$ and split each empty interval as
$P_++e^{-2\eta d}P_-$. Maximal minus strings are neutral. An all-bare
string has two walls and weight $y^2e^{-2\eta d}$; every rod has at
most two ports. A port ending in a wall costs $C|y|$, and attachment
to a new network costs $Cr_A$, using the exponential gap row sum.
The at most $4^m$ noncrossing connections of $2m$ ports sum to neutral
molecules of anchored norm $\epsilon\le C(|y|^2+r_A)$, with exponential
reserve in occupied blocks, walls and actual minus-gap lengths.
Plus holes of multi-rod tensors remain unfilled.

Neutral phases are derivative-field occupation integrals with disjoint
occupations bounded by one. Their characteristic functions have modulus
at most one, and their induced shifts satisfy the covariance row bound.
Thus the field and forest estimates of Section~\ref{s3n:extraction}
apply to occupied $E_I$ jets and $L^2$ occupation ports. They give
\begin{equation}
 e^{C|X|}C^{\sum_Id_I}\prod_Id_I!\prod_Y\|F_Y\|_{a;Y},
 \label{s3:network-block-cauchy}
\end{equation}
for $d_I$ field derivatives. Covariance edges have summable nuclear
rows; occupation ports have exponential branching, with no regulator
on empty gaps. The rooted sum of Section~\ref{s3n:extraction} therefore
bounds connected Gaussian molecules by $C\epsilon$, with size reserve left.

Gaps between distinct molecules carry $P_+$. The identity
$P_+AP_+=\langle+|A|+\rangle P_+$ factorizes their chronological evaluation,
also for interleaved rods by tensor multilinearity. The resulting
scalar complex activities $\mu_T$ retain internal compatibility.
Their symmetric pair factor $c=\prod c_{\alpha\beta}\in[0,1]$ includes
hard exclusion and soft collars.
Put $\zeta=c-1$ and $a(\gamma)=c_*s(\gamma)$, where $s$ counts
occupied blocks, walls and minus-gap length, and $c_*>0$ lies inside
the remaining reserve. Since $|1-c|\le\sum(1-c_{\alpha\beta})$, the
incidence bound gives
\[
 \int|\zeta(\gamma,\gamma')|e^{a(\gamma')}\,d|\mu_T|(\gamma')
 \le C\epsilon s(\gamma)\le a(\gamma).
\]
Also $\int e^a\,d|\mu_T|<\infty$ in finite volume. Ueltschi's
Theorem~1~\cite{Ueltschi} now gives the absolutely convergent logarithm.
The wrapping sector is bounded by $e^{-(2\eta_0-C\epsilon)T}$.
Truncation of lengths and edge ranges followed by anchored dominated
convergence gives the thermodynamic limit and holomorphy without a
first spatial moment.

This gives $|F|\le C(|y|^2+r_A)$. Invariance under
$\sigma_z:(y,R_1)\mapsto(-y,-R_1)$ gives the quadratic charged bound
by Taylor's formula. Cauchy estimates give the source and current
derivatives, including the scalar.
\end{proof}

For the complete curvature, if
$Z(h)=e^{A(h)}\int e^{f(h)M+Q_h}\,d\mu$ with $\mu$ fixed, then
\begin{equation}
 (\log Z)''=A''+\operatorname{Cov}(f'M+\partial_hQ_h,f'M+\partial_hQ_h)
                   +\E[f''M+\partial_h^2Q_h].
 \label{s3:complete-curvature}
\end{equation}
It is algebraic on a nonzero holomorphic domain; polarization supplies
spatial contacts. The source theorem retains every term.

\subsection{Calibration and selection of the physical branch}

\begin{lemma}[Positive-source calibration]
\label{s3:lem:positive-secant-calibration}
Suppose the exact fixed-depth identity is
\begin{equation}
 P_\beta(h_2)-P_\beta(h_1)
 =H_j^{-1}[\widehat P_j(H_jZ_jh_2)-\widehat P_j(H_jZ_jh_1)].
 \label{s3:secant-pressure-pullback}
\end{equation}
If $\widehat P_j(t)=c_j^0+t+O(d_j)$ on a fixed positive interval,
then $m_\beta(h)=Z_j[1+O(d_j)]$ on its compact interior. If
$Z_j\to Z_\infty>0$ and $d_j\to0$, then $m_\beta=Z_\infty$.
\end{lemma}
\begin{proof}
Convexity bounds either one-sided derivative by secants at $t\pm d$,
with fixed $d>0$. Their slopes are
$Z_j[1+O(d_j/d)]$; only $c_j^0$ cancels. Then use
$h=t/(H_jZ_j)\downarrow0$. Fixed-depth endpoint bounds justify the
thermodynamic identity before taking the depth limit.
\end{proof}

\begin{proposition}[Physical activity, trapping and thermal crossing]
\label{s3:lem:physical-charge-cone}
\label{s3:prop:nonzero-activity}
\label{s3:prop:physical-critical-trapping}
\label{s3:prop:physical-transversality}
\label{s3:prop:ordered-side-selection}
After fixed preparation, the positive-source error
$\epsilon_j^{\rm src}\le Cv_j^2+C_J\theta^j$ has arbitrarily small
supremum. Every critical weak prefix satisfies
$q_j\ge-C\epsilon_j^{\rm src}$. The physical critical entry has
nonzero reduced activity, positive in a fixed spinor gauge. It lies on the zero-limit
separator, and the physical crossing
$d_0=\partial_\beta[u_0-\Gamma_0(v_0)]_{\beta_c}$ is strictly positive.
For nearby $\beta>\beta_c$,
\begin{equation}
 u_\infty=b(2\beta Bm_\beta^2-1)
 \ge b(\beta/\beta_c-1),\qquad u_\infty\to0\quad(\beta\downarrow\beta_c).
 \label{s3:ordered-limit-charge}
\end{equation}
\end{proposition}
\begin{proof}
The source theorem and reduced-state orders bound the error.
Calibration gives $m_c\le Z_j(1+C\epsilon_j^{\rm src})$;
the jump and Ward yield $q_j\ge-C\epsilon_j^{\rm src}$.

If $v_0=0$, zero-wall invariance and stable tracking give
$Z_j=Z_\infty+O_J(\theta^j)$ and response $1+O_J(\theta^j)$.
Calibration gives $Z_\infty=m_c$. Stop at
$H_jZ_jh$ in a fixed positive interval of multiplicative width $>2L$;
then $m_c(h)-m_c=O_J(h^\alpha)$, $\alpha>0$. This contradicts the
independent bound
$m_c(h)-m_c\ge c_\varepsilon h^\varepsilon$ for every
$\varepsilon>0$ \cite[\S S7]{Jump}.
The gauge $\operatorname{diag}(1,-1)$ changes the activity sign while
preserving sources, Ward and the transformed free vector with its scalar.

Choose an outer radius $r$ and then fixed
$N$ larger than the cone constant divided by $k_-$; preparation
errors and initial charge are much smaller than $r^2$. If activity first
reaches $r/4$, the cone and charge monotonicity give $|u|=O(r^2)$.
For the next $N$ steps the map keeps $r/8<v<r/2$, while charge decreases
by $Nk_-r^2/64$, contradicting the cone. The cone and monotonicity
also exclude a charge exit. Thus the orbit stays weak,
$\sum v_j^2<\infty$, and $u_\infty\ge0$. If its limit were positive,
activity would decay geometrically; calibration would give
$Z_\infty=m_c$, contradicting Ward and $q_\infty=u_\infty>0$.
The physical orbit is therefore the zero-limit survivor.

If $d_0=0$, the stable tangent of
\eqref{s3:stable-riccati} has $T_u,T_v\to0$, bounded $T_w$.
Differentiated reconstruction gives $\partial_\beta p_j\to0$ and
$\partial_\beta(y_j,R_j)\to0$ in absolute coordinates, including
source scalars. The full base has zero temperature derivative.
At fixed admitted $\eta$, the exact response identity is
\begin{equation}
 m_\beta(\eta/(H_jZ_j))=Z_j\mathcal M_j(X_j;\eta),\quad
 \mathcal M_j=\partial_\eta\widehat P_j.
 \label{s3:positive-source-magnetization}
\end{equation}
The source theorem gives $\partial_\beta\mathcal M_j\to0$;
Ward gives $\partial_\beta Z_j\to-m_c/(2\beta_c)<0$.
Thus the field increases with $\beta$ but the right side decreases,
contradicting positive-field GKS.

For $\beta\ge\beta_c$, GKS strengthens the cone to
$q_j\ge b(\beta/\beta_c-1)-C\epsilon_j^{\rm src}$.
Trapping and calibration give \eqref{s3:ordered-limit-charge}. Positive-limit
survivors lie above the separator, so $d_0\ge0$; the contradiction
excludes zero. Finally $u_\infty(\beta)\le u_N(\beta)$ at every fixed
$N$; continuity followed by $N\to\infty$ gives the last limit.
\end{proof}

\begin{theorem}[Critical field and complete curvature]
\label{s3:thm:critical-field-profile}
At $\beta_c$, stop at $\eta=H_jZ_jh$ in a fixed positive interval. Then
\begin{equation}
 m_c(h)=Z_j+O_J(v_j^2+\theta^j)
       =m_c[1+q_j/(2b)+o(q_j)],\qquad j=\log(1/h)/b+O_J(1).
 \label{s3:actual-critical-field}
\end{equation}
Moreover
\begin{equation}
 0\le h m_c'(h)\le C_J(v_j^2+\theta^j)=o(q_j).
 \label{con:critical-differential-bound}
\end{equation}
\end{theorem}
\begin{proof}
Parity and the source theorem give response
$1+O_J(v_j^2+\theta^j)$ and curvature $O_J(v_j^2+\theta^j)$.
Use \eqref{s3:positive-source-magnetization},
$Z_j=m_c\sqrt{1+q_j/b}$ and \eqref{s2:critical-signal}.
At this fixed depth,
$h m_c'(h)=\eta Z_j\partial_\eta^2\widehat P_j\ge0$.
Neighboring charge ratios give slow variation and subpower decay;
the exact pressure integral gives the pressure equivalent in
Section~\ref{sec:critical-spatial}.
\end{proof}

\section{Microscopic spins and the common critical profile}
\label{sec:critical-spatial}

Put $j(R)=\max\{j:H_j\le R\}$ and $\epsilon_j=v_j^2+\theta^j$.
The critical signal gives $\sum_jv_j^2<\infty$, $v_j^2=o(q_j)$,
$\theta^j=o(v_j^m)$ for every $m>0$, and $v_{j+1}/v_j\to1$.
Consequently, for $\delta,m>0$ and $0<\rho<1$,
\begin{equation}
 \frac{v_i}{v_j}\le C_{J,\delta}e^{\delta|i-j|},\qquad
 \sum_{i\ge0}\rho^{|i-j|}v_i^m\le C_{J,m,\rho}v_j^m.
 \label{concise:eq:geometric-convolution}
\end{equation}
Multiply consecutive ratios and choose $\rho e^{m\delta}<1$.

\subsection{Field and pressure}
\label{s4:quantitative-critical}
\begin{proposition}[Quantitative source reconstruction]
\label{s4:prop:quantitative-critical}
For $n=j(1/h)$, $F(h)=m_{\beta_c}(h)-m_c$ and
$\mathcal P(h)=P_c(h)-P_c(0)-m_ch$,
\begin{align}
 F(h)&=m_c\bigl(\sqrt{1+q_n/b}-1\bigr)+O_J(\epsilon_n),
 \label{s4:field-exact-charge-error}\\
 \mathcal P(h)&=hm_c\bigl(\sqrt{1+q_n/b}-1\bigr)+O_J(h\epsilon_n),
 \label{s4:pressure-exact-charge-error}\\
 0\le hF(h)-\mathcal P(h)&\le C_Jh\epsilon_n.
 \label{s4:pressure-field-gap}
\end{align}
\end{proposition}
\begin{proof}
The source stop is $n+O_J(1)$. Source reconstruction, Ward and charge
drift give the first formula. The complete curvature bound
\eqref{con:critical-differential-bound} and
\eqref{concise:eq:geometric-convolution} give
$0\le\int_0^htF'(t)dt\le C_Jh\epsilon_n$.
Integrating by parts on $[\delta,h]$ and letting $\delta\downarrow0$,
using $F(\delta)\to0$ and $\mathcal P(h)=\int_0^hF(t)dt$, proves
the other two formulas.
\end{proof}

\subsection{Marked transport and spatial reconstruction}
\label{s4:rate-free-roots}\label{s4new:joint-root-transport}
\label{s4t:pinned-positive-reconstruction}

Differentiate microscopic marks at entry and use the pinned map of
Theorem~\ref{s3n:thm:common-map}. Parity and singleton closure give
$\|R_{0,j}\|\le C_J\epsilon_j$,
$\|R_{1,j}\|\le C_J(v_j^3+\theta^j)$.

\begin{lemma}[Marked transport and evaluation]
\label{s4t:prop:pinned-positive-evaluation}
\label{s4t:cor:one-spin-calibration}
Let $Z_{j,x}$ be a microscopic mark's reference-spin coefficient and
$c_{j,x},d_{j,x}$ its charged and neutral proper norms. Uniformly in $x$,
\begin{equation}
 |Z_{j,x}|\le C_J,\quad c_{j,x}\le C_Jv_j,\quad
 d_{j,x}\le C_Jv_j^2,\quad Z_{j,x}=Z_j+O_J(\epsilon_j).
 \label{s4t:eq:one-spin-calibration}
\end{equation}
For distinct sites at separation $r$, the connected joint module,
including its neutral scalar, has $C_k^{12}=O_J(v_k)$ and
$D_k^{12}=O_J(v_k^2)$ at $H_k\le r/4<H_{k+1}$.
On a fixed reserved domain $\Re\eta\ge\eta_0>0$,
\begin{equation}
 |\mathcal L_{j,\eta}(E_0+E_1)|
 \le C\{\|E_0\|+(|y_j|+\|R_{1,j}\|)\|E_1\|\}.
 \label{s4t:eq:pinned-positive-evaluation}
\end{equation}
The free reference spin evaluates to
$1+O(\|R_{0,j}\|+(|y_j|+\|R_{1,j}\|)^2)$.
\label{s4t:eq:pinned-reference-spin}
\end{lemma}
\begin{proof}
Deleting bulk root placement changes
$L^{2-m-N_q\Re p/2-D}$ to $L^{1-m-N_q\Re p/2-D}$, also inside
a close fusion group; offsets, occupied weights and currents are
unchanged. A local charged root gains $L^{-\Re p/2}$; a neutral
proper root has a current or source degree and gains $C/L$.
The pinned gap and parity therefore give
\[
 \begin{split}
 c_{j+1}&\le\kappa c_j+C(v_j+\theta^j)(|Z_{j,x}|+d_j),\\
 d_{j+1}&\le\kappa d_j+C(v_j+\theta^j)c_j+C\epsilon_j|Z_{j,x}|,\\
 |Z_{j+1,x}-Z_{j,x}|&\le C(d_j+v_jc_j+\epsilon_j|Z_{j,x}|),\qquad\kappa<1.
 \end{split}
\]
For a fixed small $\delta>0$,
$X_j=c_j/v_j+\delta d_j/v_j^2$ obeys
$X_{j+1}\le\rho X_j+C|Z_{j,x}|$, $\rho<1$, and
$|Z_{j+1,x}|\le(1+Cv_j^2)|Z_{j,x}|+Cv_j^2X_j$.
Successive maxima and $\prod_j(1+Cv_j^2)<\infty$ bound both quantities.

For two roots work in
$\mathbb C[\varepsilon_1,\varepsilon_2]/(\varepsilon_1^2,\varepsilon_2^2)$,
separating products of single-root modules from connected histories.
At the first connecting scale $i$, a path of $n$ links has one link
of length at least $r/(nH_i)$. With $w(t)=(1+t^2)^{-1}$,
\[
 \prod_{e\in\gamma}w(d_e)\le
 Cn^2w(r/H_i)\sum_{e\in\gamma}\prod_{f\ne e}w(d_f).
\]
Cut that link in the majorant. Two pinned trees remain, and the
constituent reserve sums the polynomial in $n$. Closed blobs use their
anchored second shape moment; occupied rods use their length weight.
Keeping the connecting port, scalars, contacts and point labels, parity
bounds charged and neutral births by $C_Jv_iw(r/H_i)$ and
$C_Jv_i^2w(r/H_i)$. At zero walls there is no birth; overlapping norm
covers do not identify distinct microscopic sites.

The scalar and proper zero-wall map is
$\left(\begin{smallmatrix}1&\Lambda_i^{\rm m}\\0&A_i\end{smallmatrix}\right)$,
with $\|A_i\|\le\kappa<1$ and bounded covector $\Lambda_i^{\rm m}$.
The bounded terminal shear
$S_i=\Lambda_i^{\rm m}+S_{i+1}A_i$, $S_k=0$, diagonalizes it.
The actual perturbation is $O_J(v_i)$, so
\[
 \|U_{k,i}\|\le C_Je^{C_J\sum_{l=i}^{k-1}v_l},\qquad
 \|P_NU_{k,i}P_C\|\le
 C_Je^{C_J\sum_{l=i}^{k-1}v_l}\sum_{l=i}^{k-1}v_l.
\]
The second estimate requires an odd background factor.
Since $w(r/H_i)\le CL^{-2(k-i)}$, the ratio bound in
\eqref{concise:eq:geometric-convolution}, with $3\delta<2\log L$,
sums births and conversions to $O_J(v_k)$ and $O_J(v_k^2)$.
Finite-entry terms are $O(e^{\delta k}L^{-2k})$.
The scalar component has not been contracted.

For evaluation, insert $\varepsilon E$ in the complete partition
algebra and root Theorem~\ref{s3:thm:network-positive-response} there.
The root occurs once; other incidences are small. Minus gaps retain
$e^{-2\eta_0d}$ and its first moment. The Gaussian, regulator and
occupied-support bounds give
$|[\varepsilon]\log Z|\le C\|E\|$.
\label{s4t:eq:pinned-absolute-cluster}
Differentiate source scalars before cancellation. Charged parity makes
the odd functional vanish at $(y,R_1)=0$; Cauchy gives its factor
$|y|+\|R_1\|$. The reference spin equals one there and is even,
giving its quadratic error. Integrate mass-one profiles after the
pointwise estimate. Anchored tails and the exponentially suppressed
full-circle minus sector justify the volume limit. At
$h=\eta/(H_jZ_j)$, marked and uniform-source evaluation give respectively
$m_{\beta_c}(h)=Z_{j,x}+O_J(\epsilon_j)$ and
$m_{\beta_c}(h)=Z_j+O_J(\epsilon_j)$, proving calibration.
\end{proof}

\begin{lemma}[Local state selection]
\label{lem:ferromagnetic-local-state}\label{s4t:lem:positive-local-state}
For summable ferromagnetic interactions, regular positive-field Gibbs
states tend locally to the zero-field plus state as $h\downarrow0$.
Marked periodic identities have this limit after volume tends to infinity.
\end{lemma}
\begin{proof}
At pressure differentiability, extremal states have equal one-site
means. Their monotone coupling~\cite{FKG} has
$\Pr(\sigma_x^+\ne\sigma_x^-)=(m_x^+-m_x^-)/2=0$ at every site,
so they coincide. Summability makes periodic local limits Gibbs.
Maximal states at regular $h\downarrow0$ decrease to a zero-field
Gibbs state above the zero-field maximal state; maximality gives equality.
\end{proof}

\begin{theorem}[Integer-distance reconstruction]
\label{s4new:thm:critical-point-reconstruction}
For every $J\in\mathcal J_B$,
\[
 \sup_{\substack{r\in\mathbb N\\H_j\le r<H_{j+1}}}
 |\langle\sigma_0\sigma_r\rangle_{\beta_c}^+-Z_j^2|
 \le C_Jv_j^2=o(q_j).
\]
\end{theorem}
\begin{proof}
\label{s4new:identity-fusion}\label{s4new:lem:identity-fusion}
At depth $\ell=k+O(1)$ the supports fit in a bounded window.
Regroup the full $\varepsilon_1\varepsilon_2$ numerator coefficient,
including products of single-root families, as one root retaining
both ordered point labels. The identity
$SB_y(z_2,z_1)S=B_{-y}(z_2,z_1)$ and its bounded-interval Dyson series
give charged error $O(y_\ell)$ and neutral error $O(y_\ell^2)$ after
restoring exterior transfer. The preceding lemma bounds other families.

Let $\mathbf1_i$ denote the complete insertion
$(1+\varepsilon)Z_\Lambda[\boldsymbol h]$; exact maps preserve it.
The rooted incidence bound replaces a free identity by $\mathbf1_i$
at even cost $O(\|R_0\|+|y|\|R_1\|)$ and odd cost
$O(\|R_1\|+|y|\|R_0\|)$. Extracting only its constant scalar gives
\[
 E_\ell^{12}=A_\ell\mathbf1_\ell+P_\ell,\quad
 A_\ell=Z_{\ell,0}Z_{\ell,r}+O_J(v_\ell^2),\quad
 c_\ell=O_J(v_\ell),\quad d_\ell=O_J(v_\ell^2).
\]
Other source scalars, current jets and remote constituents stay proper;
bounded internal offsets use the same finite-mark space and pinned gap.

\label{s4new:lem:homogeneous-root}
Unit invariance removes forcing proportional to $A_i$. Thus
\[
 c_{i+1}\le\kappa c_i+C_Jv_id_i,\quad
 d_{i+1}\le\kappa d_i+C_Jv_ic_i,\quad
 |A_{i+1}-A_i|\le C_J(d_i+v_ic_i).
\]
For late $\ell$, $c_i+d_i$ contracts by $\bar\kappa<1$.
The geometric convolution bound gives
$c_{\ell+n}\le C\rho_1^nv_\ell$,
$d_{\ell+n}\le C\rho_2^nv_\ell^2$, $\rho_1,\rho_2<1$, hence
$A_\infty=A_\ell+O_J(v_\ell^2)$.
Choose real $\eta$ in the reserved interval so every
$h_n=\eta/(H_nZ_n)$ is a pressure differentiability point; the excluded
values form a countable set. For the fixed pair, take volume to infinity
at $h_n$, then $n\to\infty$.
Proper evaluation vanishes and the unit evaluates to one. Local state
selection gives the plus pair as $A_\infty=Z_\ell^2+O_J(v_\ell^2)$.
Since $\ell-j(r)=O(1)$ and Ward gives
$|Z_{i+1}^2-Z_i^2|\le C_J\epsilon_i$, the band bound follows.
Separation tends to infinity after the fixed-pair source limit.
\end{proof}

\subsection{Critical profile and collective observables}
\label{con:commonprofile}\label{s4new:cor:common-critical-profile}
\begin{corollary}
\label{s4new:cor:quadratic-common-profile}\label{new:main-critical}
The physical $S(R)>0$ decreases to zero. Put $\mathcal Q(R)=q_{j(R)}$.
For fixed $a,\alpha>0$,
\begin{equation}
 \frac{\mathcal Q(aR)}{\mathcal Q(R)}\to1,\qquad
 R^\alpha\mathcal Q(R)\to\infty,\qquad S(R)\sim\mathcal Q(R)/b.
 \label{con:slowvariation}
\end{equation}
Through integer distances,
\begin{align}
 G_c(r)&=m_{\beta_c}(a/r)^2-m_c^2+O_{J,a}(v_{j(r)}^2)
 \label{s4new:eq:squared-field-spatial}\\
 &=m_c^2[S(r)+S(r)^2/4]+O_J(v_{j(r)}^2),
 \label{s4new:eq:quadratic-common-profile}\\
 G_c(r)&\sim m_c^2S(r),\quad
 m_{\beta_c}(a/r)-m_c\sim m_c\mathcal Q(r)/(2b).
 \label{con:fieldspinratio}
\end{align}
Moreover,
\begin{align}
 \mathcal P(h)&\sim hF(h)\sim m_ch\mathcal Q(1/h)/(2b),
 \label{con:pressurecommon}\\
 F(h)/h&\sim m_c\mathcal Q(1/h)/(2bh)\to\infty,
 \label{con:secant-divergence}\\
 \Var_c^+(\bar\sigma_N)&=\frac{m_c^2}{b}q_{j(N)}+O_J(v_{j(N)}^2),\notag\\
 \frac{\chi_c^{\rm tr}(N)}{2N}&=\frac{m_c^2}{b}q_{j(N)}+O_J(v_{j(N)}^2).
 \label{s4new:eq:collective-activity-errors}
\end{align}
The quadratic term need not exceed the activity error.
\end{corollary}
\begin{proof}
Consecutive slow variation of $q_j$ gives the first limit and
$\log q_j=o(j)$, hence the power comparison. The source stop gives
$m_{\beta_c}(a/r)=Z_{j(r)}+O_J(v_{j(r)}^2)$.
Squaring and applying the band theorem, Ward and the definition of $S$
proves the spatial formulas. The source proposition gives the field,
pressure and secant formulas; field monotonicity gives monotonicity of $S$.

For the collective errors choose $1<\lambda<L$. At late $j\ge i$,
$v_i^2\le C_J\lambda^{j-i}v_j^2$ and
$|q_i-q_j|\le C_Jv_j^2\sum_{d=1}^{j-i}\lambda^d$.
The fraction of $r\le N$ in band $i$ is $O_L(L^{i-j(N)})$;
geometric summation gives error $O_J(v_{j(N)}^2)$.
Insert $G_c(r)=m_c^2q_{j(r)}/b+O_J(v_{j(r)}^2)$ in the exact
susceptibility and variance sums, with weights $1/N$ and
$2(1-r/N)/N$. The latter sum to $1-1/N$.
The finite prefix and diagonal contribute
$O_J(N^{-1})=o(v_{j(N)}^2)$.
\end{proof}

\section{Thermal classification and logarithmic tails}
\label{sec:thermal}

All maps act on the fixed physical temperature curve. Write
$w=\log v$, $\gamma_j(w)=\Gamma_j(e^w)$, and
$\ell_j=\gamma_j'(w_j^c)\ge0$. Let $\Pi_j$ be the complete normal
product and $d_0>0$ the transverse entry derivative.
Superscripts $c,+,-$ denote the critical orbit and detunings
$\beta_c\pm\tau$; primed state variables denote one step.

\subsection{The ordered endpoint}
\label{s4:gain-limit}

\begin{lemma}[Discrete Volterra estimate]
\label{concise:lem:volterra}
Let $0\le M<T\le\infty$, $\rho_j\ge0$ for $j\ge M$, and
$d_j,z_j\ge0$ for $M\le j\le T$. Suppose, for $M\le j<T$,
\[
 |d_{j+1}-d_j|\le C\rho_j(d_j+z_j),\qquad
 z_{j+1}\le(1+C\rho_j)z_j+Cd_j.
\]
If $\sum_{j\ge M}(j+1)\rho_j\le1$ and
$\eta=d_M+z_M/(M+1)$, then, uniformly in $M,T$,
\begin{equation}
 \begin{split}
 d_j&\le C_1\eta,\qquad z_j\le C_1\eta(j+1),\\
 |d_j-d_N|&\le C_1\eta\sum_{i=N}^{j-1}(i+1)\rho_i,
 \qquad M\le N\le j\le T.
 \end{split}
 \label{concise:eq:volterra}
\end{equation}
Here $C_1$ depends only on $C$.
\end{lemma}
\begin{proof}
With $D_j=\max_{M\le i\le j}d_i$, variation of constants gives
$z_j\le C_2[z_M+(j-M)D_j]$. Substitution in the first inequality gives
$D_j\le C_3\eta+C_3\sum_{i=M}^{j-1}(i+1)\rho_iD_i$.
Iteration, equivalently discrete Gronwall~\cite{Clark1987}, bounds this by
$C_3\eta\prod_{i=M}^{j-1}[1+C_3(i+1)\rho_i]\le C_3e^{C_3}\eta$.
Substitute back and sum the first inequality.
\end{proof}

The tangential recurrence following \eqref{s3:stable-riccati} gives
$|\log v_j^c(w)-\log v_j^c(w_c)|\le C|w-w_c|$ uniformly in $j$.
Thus the moment tails below are uniformly dominated near $w_c$.

\begin{theorem}[Physical slope criterion]
\label{s4:prop:gain-limit}\label{s4t:thm:gain-endpoint-dichotomy}
\label{s4t:cor:physical-field-cusp-test}\label{new:main-thermal}
The following conditions are equivalent:
\begin{equation}
 \sum_j(j+1)(v_j^c)^2<\infty
 \ \Longleftrightarrow\ \sum_j\ell_j<\infty
 \ \Longleftrightarrow\ \Pi_\infty:=\lim_j\Pi_j<\infty
 \ \Longleftrightarrow\
 \int^\infty S(R)\frac{dR}{R}<\infty.
 \label{s4t:eq:moment-gain-equivalence}
\end{equation}
The integral may be replaced by
$\int_0[m_{\beta_c}(h)-m_c]\,dh/h$ or $\sum_{r\ge1}G_c(r)/r$.
In the finite case $\kappa_J=d_0\Pi_\infty/b>\beta_c^{-1}$ and
\begin{equation}
 E_+(\tau)=\kappa_J\tau+o(\tau),\qquad
 m_{\beta_c+\tau}-m_c
 =\frac{m_c}{2}(\kappa_J-\beta_c^{-1})\tau+o(\tau).
 \label{s4t:eq:finite-gain-cusp}
\end{equation}
Otherwise $E_+(\tau)/\tau\to\infty$ and
$m_{\beta_c+\tau}-m_c\sim m_cE_+(\tau)/2$.
The right ordered slope therefore exists in $(0,\infty]$.
\label{s4t:eq:infinite-gain-cusp}
\label{s4t:eq:right-temperature-derivative}
\end{theorem}
\begin{proof}
Put $r_j=(v_j^c)^2$. Since $\sum r_j<\infty$,
$\log M_j=\log[1+(1+a_j/b)\ell_{j+1}]+O(r_j)$.
The leading term is eventually nonnegative and comparable to
$\ell_{j+1}$, proving the product criterion.
For the tangent $(U_j,-W_j)$ initially $(1,0)$, divide out the
convergent positive diagonal products. Its positive recurrence becomes
$\Delta\widetilde U_j=b_j\widetilde W_j$,
$\Delta\widetilde W_j=c_j\widetilde U_j$, where $b_j\asymp r_j$,
$c_j\asymp1$. Thus $\widetilde W_j\ge cj$ and $U_j$ is unbounded
when $\sum(j+1)r_j=\infty$. Conversely, the affine Riccati estimate
in \eqref{s3:stable-riccati} gives
\begin{equation}
 0\le\ell_j\le C\sum_{k\ge j}r_k.
 \label{s4t:eq:riccati-tail-charge}
\end{equation}
A finite moment therefore gives $\sum\ell_j<\infty$.
The identity $U_j+\ell_jW_j=\Pi_j$ completes the equivalence and, in
the finite case, gives $W_j=O(j)$, $j\ell_j\to0$ and $U_j\to\Pi_\infty$.
\label{s4t:eq:finite-tangent-endpoint}

Write the ordered physical entry as $(\gamma_0(w)+\delta,w)$, with
$\delta=d_0\tau+o(\tau)$ and $w=w_c+O(\tau)$, and compare it with the
critical point at the same $w$. Then $d_j=u_j^+-u_j^c\ge0$ and
$z_j=\log(v_j^c/v_j^+)\ge0$. The mean-value matrix satisfies the lemma
with $\rho_j=r_j$, even for unbounded $z_j$, since intervening
activities are at most $v_j^c$. When the moment is finite, the lemma
gives, after a fixed prefix,
\[
 d_j\le C\delta,\quad z_j\le C\delta(j+1),\qquad
 |u_\infty(\delta,w)-d_N|
 \le C\delta\sum_{j\ge N}(j+1)(v_j^c(w))^2.
\]
Uniform tails permit finite-depth differentiation followed by
$N\to\infty$: $u_\infty/\delta\to\Pi_\infty$.
If the moment diverges, positivity instead gives
$u_\infty\ge d_N\prod_{j\ge N}(1-Cr_j)$.
Take $\tau\downarrow0$ at fixed $N$, where $d_N/\tau\to d_0U_N$
by differentiation of the finite iterate. The tail product tends to one
as $N\to\infty$ because the squared activities are summable. Since the
entry derivative is positive, $U_N\to\infty$ forces
$u_\infty/\tau\to\infty$.
At a later entry scale the initial derivative absorbs the prefix
$\Pi_N$ of the normal product. The gain from that scale onward equals
the full product divided by this prefix, giving the same endpoint
derivative.

Calibration gives $q_\infty=u_\infty=bE_+$, and Ward gives
$m_{\beta_c+\tau}=m_c\sqrt{(1+E_+)/(1+\tau/\beta_c)}$.
For strict positivity choose $J_{n_*}>0$. In periodic volumes, the
one-site conditional law bounds the disagreement probability below by
$p_*=[1+\exp(2h_1+4\beta_1\sum_nJ_n)]^{-1}$ on a compact temperature
interval and $0<h\le h_1$. The two incident bonds and GKS give
$\partial_\beta m_{\beta,V}(h)\ge4p_*J_{n_*}m_{\beta,V}(h)$.
Integrating before the volume and $h\downarrow0$ limits yields
\begin{equation}
 m_{\beta_c+\tau}\ge m_ce^{\alpha_J\tau},\qquad
 \alpha_J=4p_*J_{n_*}>0,
 \label{s4t:eq:strict-cusp-lower}
\end{equation}
hence $\kappa_J>\beta_c^{-1}$.

Finally $\sum_j u_j^c=\sum_k(k+1)k_k(v_k^c)^2$.
Use $q_j^c\sim u_j^c$ and $S(R)\sim q_j^c/b$ on logarithmic bands
of width $b$. The field substitution is exact; the spatial test follows
from $G_c(r)\sim m_c^2S(r)$ and monotonicity of $S$.
\label{new:spatial-slope-criterion}\label{new:spatial-slope-equivalence}
\end{proof}

\subsection{Exit and susceptibility}
\label{s4t:general-continuation}\label{s4o:finite-curvature-section}

\begin{proposition}[Exit and physical readout]
\label{s4t:prop:fixed-outgoing}\label{s4o:prop:finite-curvature}
\label{s4o:lem:superlog-exit}\label{s4o:prop:exit-lower}
For sufficiently small fixed $\varepsilon>0$ and late fixed
$J_\varepsilon$, the disordered orbit reaches
$N_\varepsilon(\tau)=\min\{j\ge J_\varepsilon:v_j^-\ge\varepsilon\}<\infty$,
remaining in the weak charts through this index. Writing $N=N_\varepsilon$,
\begin{equation}
 \begin{gathered}
 v_N=\varepsilon+O(\varepsilon^2),\quad
 u_N=-\varepsilon+O(\varepsilon^2),\quad
 q_N=u_N+O(\varepsilon^2)+o_\tau(1),\\
 N_\varepsilon(\tau)/\log(1/\tau)\to\infty.
 \end{gathered}
 \label{s4t:eq:fixed-outgoing-state}
\end{equation}
Changing between fixed small sections $\varepsilon_1,\varepsilon_2$
changes the stopping time by $O_{J,\varepsilon_1,\varepsilon_2}(1)$.
The complete terminal source pressure satisfies
\begin{align}
 \widehat\chi_N=\widehat P_N''(0)&=
          \frac{\chi_-(\beta_c-\tau)}{H_NZ_N^2},
 \label{s4o:eq:complete-fixed-depth-curvature}\\
 \log\chi_-(\beta_c-\tau)&=bN+\Delta_-(\tau),\qquad
 \Delta_-(\tau)\ge-C_{J,\varepsilon}.
 \label{new:thermal-residual}
\end{align}
Here $\Delta_-=\log(H_0Z_N^2\widehat\chi_N)$ retains the complete
terminal curvature. In particular $\log\chi_-/\log(1/\tau)\to\infty$.
No unrestricted upper estimate is asserted.
\label{s4o:eq:superalgebraic-susceptibility}
\end{proposition}
\begin{proof}
Choose $J_\varepsilon$ so that future $|a_j|\le\varepsilon^2$,
$\bar\sigma_j\le\varepsilon$, and critical coordinates are
$O(\varepsilon^2)$. Fixed-depth continuity puts the detuned point there;
separator classification makes its decreasing charge eventually negative.
On $|u|\le2\varepsilon$, $v\le\varepsilon$,
\[
 |\Delta(u^2-v^2)|
 \le C(u^2+v^2+|u|\bar\sigma_j+|a_j|)v^2
 \le C\varepsilon^2v^2.
\]
Charge telescoping gives $\sum v^2\le C\varepsilon$, hence
$|u^2-v^2|\le C\varepsilon^3$, excluding charge exit at $-2\varepsilon$.
A negative-charge orbit remaining in this box would have both
$v_j\to0$ and eventual geometric growth. Thus activity exits, with
the stated overshoot and charge. Between fixed sections $r$ and $2r$,
eventually $\log(v'/v)\ge r/4$, proving bounded section change.
\label{s4t:eq:exit-invariant}\label{s4t:eq:exit-choice}
For each $\delta>0$ a fixed late small ball in $(u,v)$ coordinates
has $\|DF_j\|\le e^\delta$.
The physical displacement remains $O_\delta(\sqrt\tau)$ through
$j\le\log(1/\tau)/(2\delta)$, closing ball admissibility.
Let $\delta\downarrow0$.

Sharpness~\cite[Theorem~2.1(2)]{DT} gives $\chi_-(\beta)<\infty$.
GKS and GHS~\cite{GKS,Lebowitz} bound free-volume covariances at $h\ge0$
by summable infinite-volume zero-field correlations.
They give $m_\beta(h)\le h\chi_-(\beta)$; restriction to a fixed volume
gives the reverse limit. Therefore
\[
 m_\beta(h)/h\to\chi_-(\beta),\qquad
 P_\beta(h)-P_\beta(0)=\tfrac12\chi_-(\beta)h^2+o(h^2).
 \label{s4o:eq:physical-zero-curvature}
\]
The same majorant gives the upper limit of free-volume curvatures;
restricting pairs around bulk sites to a fixed interval, then enlarging
it, gives the lower. At fixed $N$, differentiate the exact source
identity with $\eta=H_NZ_Nh$, retaining source scalars and endpoints
before the volume limit. This proves the complete-curvature identity.

At exit the positive-source estimate gives
$\mathcal M_N(\eta_*)\ge c_*>0$ at fixed admitted $\eta_*>0$:
its stable error is $O(\varepsilon)+o_\tau(1)$.
Ward bounds $Z_N$ away from zero. With $h_*=\eta_*/(H_NZ_N)$,
\[
 \chi_-\ge m_{\beta_c-\tau}(h_*)/h_*
 =H_NZ_N^2\mathcal M_N(\eta_*)/\eta_*
 \ge c_{J,\varepsilon}H_N.
\]
Taking logarithms proves the lower bound.
\end{proof}

\paragraph{Ordered endpoint estimate.}
\label{s4t:prop:ordered-certificate}
At $(u_N,v_N)=(a,z)$, $a>0$, put $A_N=\sup_{j\ge N}|a_j|$,
and let $C_h$ bound $|h_j|$ on the reserved weak domain.
If
\begin{equation}
 (1+a/b)A_N\le a/8,\quad C_hz^2\le a/8,\quad
 R_N=\frac{k_+z^2}{1-e^{-a/2}}\le a/2,
 \label{s4t:eq:ordered-certificate}
\end{equation}
then $a-R_N\le u_\infty\le a$ and $v_{N+k}\le ze^{-ak/4}$.
\label{s4t:eq:ordered-error}
Indeed $u\ge a/2$ implies this activity bound; its square sum bounds
charge loss by $R_N$ and closes the assumption. Every fixed ordered
branch eventually satisfies these conditions.

\subsection{Finite gain and the disordered window}
\label{s4t:finite-gain-examples}\label{s4o:physical-failure-section}
\begin{proposition}
\label{s4t:thm:finite-gain-exit}\label{s4o:cor:finite-gain-matching}
Under the finite-slope criterion,
\begin{equation}
 \tau N_\varepsilon(\tau)\to\infty,\qquad
 \tau\log\chi_-\to\infty,\qquad E_+(\tau)\log\chi_-\to\infty.
 \label{s4o:eq:finite-gain-physical-divergence}
\end{equation}
Uniformly for $t$ in compact subsets of $(0,\infty)$,
\begin{equation}
 u^-_{\lfloor t/\tau\rfloor}/\tau\to-b\kappa_J,\qquad
 \log(v^-_{\lfloor t/\tau\rfloor}/v^c_{\lfloor t/\tau\rfloor})
 \to b\kappa_Jt.
 \label{s4t:eq:finite-gain-window}
\end{equation}
\end{proposition}
\begin{proof}
Compare the disordered physical entry $(\gamma_0(w)-\delta,w)$ with
its critical point, where $\delta=d_0\tau+o(\tau)$ and
$w=w_c+O(\tau)$. Now $d_j=u_j^c-u_j^-\ge0$,
$z_j=\log(v_j^-/v_j^c)\ge0$.
On $z_j\le B$ the same mean-value bounds satisfy
Lemma~\ref{concise:lem:volterra} with $\rho_j=e^{2B}(v_j^c)^2$.
Fix $K$. The finite tangent bound gives
$d_M+z_M/(M+1)\le D\delta$ at each fixed $M$, with $D$ independent of $M$,
once $\delta$ is small. Choose $B>2C_1D(K+1)$, then a fixed $M$
with $e^{2B}\sum_{j\ge M}(j+1)(v_j^c)^2\le1$.
The lemma strictly improves $z\le B$ through $j\le K/\delta$.
Taking $M$ later keeps this interval below any fixed activity section;
the fixed prefix follows by continuity. Since $K$ is arbitrary,
$\tau N_\varepsilon\to\infty$.

The tail bound in \eqref{concise:eq:volterra}, first at
$\delta\downarrow0$ with fixed $M$, then $M\to\infty$, gives
$d_j/\delta\to\Pi_\infty$ on these windows.
Also $u_j^c=o(j^{-1})$ and
$z_{j+1}-z_j=(1+a_j/b)d_j+O((v_j^-)^2(d_j+z_j))$.
Summation, using $a_j\to0$ and the moment tail, gives
$z_j-\Pi_\infty\delta j\to0$.
Uniform logarithmic tangential control replaces the nearby critical
family by the physical critical orbit. This proves the window limit.
The residual lower bound and $E_+\sim\kappa_J\tau$ give the divergences.
\end{proof}

\subsection{A microscopic logarithmic tail}
\label{new:logarithmic-family}
\begin{theorem}
\label{s4t:prop:microscopic-finite-gain}
\label{s4t:thm:log-tail-stopping}
\label{s4o:cor:log-tail-physical-divergence}
Fix $c>1$ and an arbitrary finite nonnegative core, with
\begin{equation}
 J_n=\frac B{n^2}\left(1+\frac c{\log(n+e)}\right)
 \quad\hbox{outside the core}.
 \label{s4t:eq:microscopic-log-tail}
\end{equation}
For $s_j=j+\log H_0/b$, the heat construction satisfies
\begin{equation}
 a_j=\frac c{s_j}+O_J(s_j^{-2}),\quad
 v_j^c\sim V_Js_j^{-c},\quad
 u_j^c\sim q_j^c\sim\frac{V_J^2}{2c-1}s_j^{1-2c},\quad V_J>0.
 \label{s4t:eq:finite-gain-critical-profile}
\end{equation}
It has finite gain, the linear cusp above, and
\begin{align}
 S(R)&\sim\frac{V_J^2b^{2c-2}}{2c-1}(\log R)^{1-2c},
 \label{s4t:eq:finite-gain-field-profile}\\
 N_\varepsilon(\tau)&\sim
       \frac{c-1}{b\kappa_J\tau}\log(1/\tau),
 \label{s4t:eq:log-tail-stopping}\\
 \liminf_{\tau\downarrow0}\frac{\tau\log\chi_-}{\log(1/\tau)}
 &\ge\frac{c-1}{\kappa_J},\qquad
 \liminf_{\tau\downarrow0}\frac{E_+\log\chi_-}{\log(1/\tau)}
 \ge c-1.
 \label{s4o:eq:log-tail-physical-divergence}
\end{align}
The last line gives physical lower bounds.
\end{theorem}
\begin{proof}
For $x=|k|$, split the logarithmic Fourier sum at
$n=x^{-1/2},x^{-3/2}$. The exterior ranges are $O(x^{3/2})$ since
$1-\cos nx\le\min((nx)^2/2,2)$. In the middle,
replacing $1/\log(n+e)$ by $1/\log(1/x)$ costs
$C(1+|\log nx|)/\log^2(1/x)$, and
$\sum_n(1-\cos nx)n^{-2}(1+|\log nx|)\le Cx$.
Together with $2\sum_n(1-\cos nx)/n^2=\pi x-x^2/2$,
\begin{equation}
 d_J(k)=\pi|k|\left[1+\frac c{\log(1/|k|)}
                 +O(\log^{-2}(1/|k|))\right]+O_J(k^2).
 \label{s4t:eq:abelian-log-tail}
\end{equation}
Finite-core changes contribute $O_J(k^2)$.
In the heat shell put $\xi=Hk$; the kernel
$(e^{-\xi^2}-e^{-L^2\xi^2})/\xi$ has integral $b$ and finite logarithmic
moments. On $H^{-1/2}\le\xi\le H^{1/2}$, replacing
$(\log H-\log\xi)^{-1}$ by $(\log H)^{-1}$ costs
$O((1+|\log\xi|)/\log^2H)$; exterior ranges and lattice errors are
$O_J(H^{-1})$. Hence $\nu_H=b+bc/\log H+O_J(\log^{-2}H)$.

Choose $1<c'<c$. Positive critical charge gives
$\log(v_{j+1}^c/v_j^c)\le-c'/s_j+C(v_j^c)^2$, hence
$v_j^c\le Cs_j^{-c'}$ and a finite first moment.
The terms $u_j^c$, $u_j^ca_j$, $(v_j^c)^2$ and
$a_j-c\log(1+1/s_j)$ are summable. Thus the activity equation makes
$\log(v_j^cs_j^c)$ converge to a finite real value.
Backward charge summation gives the charge law; $S\sim q/b$ gives
the physical profile.

For stopping, fix large $M$ and let $j_M=\lfloor M/\tau\rfloor$.
The window limit gives
$u^-_{j_M}=-b\kappa_J\tau+o(\tau)$ and
$v^-_{j_M}=V_JM^{-c}e^{b\kappa_JM}\tau^c[1+o(1)]$.
Until $v=\eta\tau$, for fixed small $\eta>0$, the future defect lies
between zero and $2c\tau/M$. Activity grows at rate at least
$(b\kappa_J-2c/M-o(1))\tau$. Its backward square sum is
$O(\eta^2\tau)$, bounding $-u$ between
$(b\kappa_J-o(1))\tau$ and $(b\kappa_J+C\eta^2+o(1))\tau$.
The required logarithmic growth is
$(c-1)\log(1/\tau)+O_{M,\eta,J}(1)$.
The crossing time therefore has lower and upper leading coefficients
$(c-1)/(b\kappa_J+C\eta^2)$ and $(c-1)/(b\kappa_J-2c/M)$.

At that crossing $a_j=o(\tau)$. Reaching $v=b\kappa_J\tau$ takes
$O_{J,\eta}(\tau^{-1})$ steps. Thereafter $Q=-u\ge v/2$ is invariant:
at its boundary
$Q'-v'/2=(k_j-1/4)v^2+O(v^3+|a_j|v)>0$, with $k_j\ge3/4$.
Inside, $\log(v'/v)\ge v/3$, so $1/v-1/v'\ge c_*>0$.
The remaining passage takes $O_J(\tau^{-1})$ steps, and
$\sum v^2\le C\varepsilon$ keeps charge in the chart.
Let $\tau\downarrow0$, then $M\to\infty$ and $\eta\downarrow0$.
The stopping equivalent and the complete residual lower bound prove
the physical inequalities.
\end{proof}

\section{Exact-tail critical and ordered benchmarks}
\label{sec:exact-tail-benchmark}\label{sec:sharp}

Assume $J_n=B/n^2$ outside an arbitrary finite nonnegative core.
Use the fixed heat coordinates and the original detuning $\beta_c+\tau$.

\begin{theorem}[Exact-tail benchmark]
\label{focused:exact-ordered-benchmark}\label{new:exact-critical-sharp}
\label{s3:prop:canonical-critical-interface}
The critical coordinates satisfy $q_j^c\sim u_j^c\sim v_j^c\sim j^{-1}$.
For $\ell_h=\log(1/h)$, as $r\to\infty$, $h\downarrow0$,
\begin{align}
 G_c(r)&=\frac{m_c^2}{\log r}
      +O_J\!\left(\frac{\log\log r}{\log^2r}\right),\notag\\
 m_{\beta_c}(h)-m_c&=\frac{m_c}{2\ell_h}
      +O_J\!\left(\frac{\log\ell_h}{\ell_h^2}\right),\notag\\
 P_c(h)-P_c(0)-m_ch&=\frac{m_ch}{2\ell_h}
      +O_J\!\left(\frac{h\log\ell_h}{\ell_h^2}\right).
 \label{new:exact-critical-errors}
\end{align}
Both $\Var_c^+(\bar\sigma_N)$ and $\chi_c^{\rm tr}(N)/(2N)$ equal
$m_c^2/\log N+O_J(\log\log N/\log^2N)$. The intrinsic limit
\begin{equation}
 \lambda_J:=\lim_{\tau\downarrow0}\frac{E_+(\tau)^2}{\tau}
 \in(0,\infty)
 \label{focused:exact-intrinsic-metric}
\end{equation}
exists, and
\begin{equation}
 E_+(\tau)\sim\sqrt{\lambda_J\tau},\qquad
 m_{\beta_c+\tau}-m_c\sim\frac{m_c}{2}\sqrt{\lambda_J\tau}.
 \label{focused:exact-ordered-laws}
\end{equation}
\end{theorem}
\begin{proof}
The heat distance, shell defect and quadratic-memory error decay
geometrically. For $I_j=u_j^2-v_j^2$, the map gives
$|\Delta I_j|\le C(u_j^2+v_j^2+\vartheta^j)v_j^2$, $0<\vartheta<1$.
Put $\eta_j=\sup_{i\ge j}v_i^2/u_i\to0$. Backward charge summation gives
$|I_j|\le Cu_j^3+C\eta_ju_j^2+C\vartheta^ju_j=o(u_j^2)$, using $\log u_j=o(j)$. Thus $v_j/u_j\to1$ and
$\Delta(1/u_j)\to1$. Now $\Delta I_j=O(j^{-4})$, so
$I_j=O(j^{-3})$ and $u_{j+1}=u_j-u_j^2+O(j^{-3})$. Hence
\begin{equation}
 u_j^c,v_j^c,q_j^c=j^{-1}+O_J(\log j/j^2).
 \label{s4t:eq:canonical-critical-rate}
\end{equation}
The quantitative formulas of Section~\ref{sec:critical-spatial}, with
$v_j^2=O(j^{-2})$ and $bj=\log r+O_J(1)$ or $\ell_h+O_J(1)$,
give the critical conclusions.

The Riccati tail bound and its recurrence on $[j,2j]$ give
$c/j\le\ell_j\le C/j$. Substitution of $1/j$ leaves residual
$O(\log j/j^3)$; the slope derivative
$\det D(f,g)/(f_u-\ell g_u)^2\le1-c'/j$ sums it backward to
$\ell_j=j^{-1}+O(\log j/j^2)$. Therefore $M_j=1+j^{-1}+O(\log j/j^2)$ and
$\Pi_j/j=P_*+O(\log j/j)$, with $P_*\in(0,\infty)$.
With the physical $d_0>0$ of Section~\ref{sec:reduction-sources}, set
\begin{equation}
 s_\tau=(d_0P_*\tau)^{-1/2}.
 \label{focused:exact-normal-scale}
\end{equation}

For $X_j=(u_j,v_j)$,
\[
 DF_j(X_j^c)=I+j^{-1}
 \begin{pmatrix}0&-2\\-1&-1\end{pmatrix}
 +O(\log j/j^2).
\]
The eigenvalues $1,-2$ and summable remainder bound its fundamental
matrix by $\|\Phi(j,k)\|\le Cj/k$ for $j\ge k\ge1$.
For $T_j=\partial_\beta X_j|_{\beta_c}$, the exact identity
$T_{u,j}-(\ell_j/v_j^c)T_{v,j}=d_0\Pi_j$ and
$\ell_j/v_j^c=1+O(\log j/j)$ give, with
$z_j=T_{v,j}/j$,
\[
 z_{j+1}=[1-3/j+O(\log j/j^2)]z_j
       -d_0P_*/j+O(\log j/j^2).
\]
Subtract $-d_0P_*/3$ and sum the stable recurrence:
$z_j\to-d_0P_*/3$, hence $T_j=d_0P_*j(2,-1)/3+o(j)$.
Bounded map Hessians and the fixed $C^2$ entry give, with the finite
initial contribution absorbed into $C\tau^2j$,
\[
 |X_j(\beta_c+\tau)-X_j^c-\tau T_j|
 \le C\tau^2j+C\sum_{1\le k<j}(j/k)|X_k(\beta_c+\tau)-X_k^c|^2.
\]
For $\tau j^2\le\delta_*$, the bootstrap
$|X_k-X_k^c|\le C_1\tau k$ closes with error $O(\tau^2j^3)$.

Write $s=s_\tau$, $X_j^+=X_j(\beta_c+\tau)$, and let $Y_\tau$
interpolate $sX_j^+$ linearly at $t_j=j/s$. For fixed sufficiently
small $t_0>0$ and $j_0=\lfloor t_0s\rfloor$, the preceding estimates give
\[
 sX_{j_0}^+=t_0^{-1}(1,1)+\frac{t_0}{3}(2,-1)
             +O_J(t_0^3)+o_\tau(1).
\]
Proposition~\ref{s3:prop:ordered-side-selection} gives
$0<u_j^+\le u_{j_0}^+$ for $j\ge j_0$.
Charge summation and the activity equation imply
\[
 \sum_{j\ge j_0}(v_j^+)^2\le\frac{u_{j_0}^+}{k_-},\qquad
 v_j^+\le v_{j_0}^+\exp[Cu_{j_0}^++C\vartheta^{j_0}].
\]
Thus $Y_\tau$ is uniformly bounded on each $[t_0,T]$. On that interval,
the reduced map, with $F(U,V)=(-V^2,-UV)$, gives
\[
 Y_\tau(t_{j+1})=Y_\tau(t_j)+s^{-1}F(Y_\tau(t_j))
                 +O_{J,t_0,T}(s^{-2}).
\]
This also gives equicontinuity. Every subsequential limit on compact
subsets of $(0,\infty)$ solves $Y'=F(Y)$ and satisfies
$Y(t)=t^{-1}(1,1)+t(2,-1)/3+O_J(t^3)$ as $t\downarrow0$.
Since $U^2-V^2$ is constant, the entrance expansion fixes it to $2$.
Its positive solution with this singularity is
\begin{equation}
 U_+(t)=\sqrt2\coth(\sqrt2t),\qquad
 V_+(t)=\sqrt2\operatorname{csch}(\sqrt2t).
 \label{focused:exact-ordered-passage}
\end{equation}
Uniqueness identifies every subsequential limit, so
$Y_\tau\to(U_+,V_+)$ locally uniformly on $(0,\infty)$.

For fixed $T$, put $N=\lfloor Ts\rfloor$, $a=u_N^+$, $z=v_N^+$,
and $A_N=\sup_{j\ge N}|a_j|=O_J(\vartheta^N)$.
Then $sa\to U_+(T)$, $sz\to V_+(T)$, and the first two ratios
$(1+a/b)A_N/a$ and $C_hz^2/a$ tend to zero. For the charge-loss
bound $R_N$ in \eqref{s4t:eq:ordered-certificate},
$R_N/a\to2k_+V_+(T)^2/U_+(T)^2$.
Choose $T$ large enough that this limit is less than $1/2$.
The certificate then applies for all sufficiently small $\tau$, yielding
\begin{align*}
 U_+(T)-\frac{2k_+V_+(T)^2}{U_+(T)}
 &\le\liminf_{\tau\downarrow0}s u_{\infty,+}\\
 &\le\limsup_{\tau\downarrow0}s u_{\infty,+}\le U_+(T).
\end{align*}
Now $T\to\infty$ gives $s_\tau u_{\infty,+}\to\sqrt2$.
Heat calibration gives $q_{\infty,+}=u_{\infty,+}=bE_+$, hence
$\lambda_J=2d_0P_*/b^2>0$; Ward gives the magnetization law.
The limit defining $\lambda_J$ uses the ordered magnetization at the
physical detuning. Its value is therefore independent of the preparation
scale and heat coordinates.
\end{proof}

\bibliographystyle{apsrev4-2}
\makeatletter
\if@filesw\immediate\write\@auxout{\string\citation{UCRSuppStyle}}\fi
\makeatother
\bibliography{references}